\documentclass[conference]{IEEEtran}

\usepackage{cite}
\usepackage{amsmath,amssymb,amsfonts}
\usepackage{algorithmic}
\usepackage{graphicx}
\usepackage{textcomp}
\usepackage{xcolor}
\usepackage{booktabs} 
\usepackage{gensymb}
\usepackage{array}
\usepackage{chemist}
\usepackage{tcolorbox}
\usepackage{hyperref}
\usepackage[margin=0.8in]{geometry}
\usepackage{lscape}
\usepackage{adjustbox}
\usepackage{indentfirst}
\usepackage{multicol, blindtext}
\usepackage{pifont}
\usepackage{lettrine}
\usepackage{pdfpages}

\PassOptionsToPackage{hyphens}{url}
\usepackage{hyperref}

\newcommand{\coo}{CO\textsubscript{2}}

\usepackage{ragged2e}

\hypersetup{%
colorlinks=true,
linkcolor=blue,
citecolor=blue,
urlcolor=blue}

\usepackage{verbatim}

\definecolor{forestgreen}{rgb}{0.13, 0.55, 0.13}
\newcommand{\cgmark}{\textcolor{forestgreen}{\checkmark} }
\newcommand{\xmark}{\textcolor{red}{\ding{55}}}

\begin{document}



\setcounter{page}{1}
\title{From Silicon Shield to Carbon Lock-in?\\The Environmental Footprint of\\Electronic Components Manufacturing\\in Taiwan (2015-2020)}

\author{\IEEEauthorblockN{Gauthier Roussilhe$^\ddagger$, Thibault Pirson$^\dagger$, Mathieu Xhonneux$^\dagger$, David Bol$^\dagger$}
\IEEEauthorblockA{
\textit{$^\ddagger$RMIT}, Royal Melbourne Institute of Technology, Australia \\
\textit{$^\dagger$ICTEAM}, Université catholique de Louvain, Belgium \\
Email: gauthierroussilhe@protonmail.com \\
}
}


\makeatletter
\patchcmd{\@maketitle}
  {\addvspace{-1\baselineskip}\egroup}
  {}
  {}
\makeatother

\maketitle

\begin{abstract}
Taiwan plans to rapidly increase its industrial production capacity of electronic components while concurrently setting policies for its ecological transition. Given that the island is responsible for the manufacturing of a significant part of worldwide electronics components, the sustainability of the Taiwanese electronics industry is therefore of critical interest. In this paper, we survey the environmental footprint of 16 Taiwanese electronic components manufacturers (ECM) using corporate sustainability responsibility reports (CSR). Based on data from 2015 to 2020, this study finds out that our sample of 16 manufacturers increased its greenhouse gases (GHG) emissions by 7.5\% per year, its final energy and electricity consumption by 8.8\% and 8.9\%,  and the water usage by 6.1\%. We show that the volume of manufactured electronic components and the environmental footprints compiled in this study are strongly correlated, which suggests that relative efficiency gains are not sufficient to curb the environmental footprint at the national scale. Given the critical nature of electronics industry for Taiwan's geopolitics and economics, the observed increase of energy consumption and the slow renewable energy roll-out, these industrial activities could create a carbon lock-in, blocking the Taiwanese government from achieving its carbon reduction goals and its sustainability policies.
Besides, the European Union, the USA or even China aim at developing an industrial ecosystem targeting sub-10nm CMOS technology nodes similar to Taiwan. This study thus provides important insights regarding the environmental implications associated with such a technology roadmap. All data and calculation models used in this study are provided as supplementary material.
\end{abstract}



\section{Introduction}

\lettrine{O}VER the last four decades, Taiwan has developed a strong ecosystem in semiconductor manufacturing, building a leadership position in packaging, foundries, and testing of integrated circuits (ICs). The island is a key provider of electronic components for the rest of the world with more than 63\% of the 2021 worldwide output value ensured by Taiwanese IC foundries and 58\% of the IC packaging \cite{itri2022}. The electronics industry accounts on average for 30\% of Taiwan's exports \cite{MOF-2021} and it is therefore not surprising that it significantly contributes to the economy of the island \cite{chou2019conundrums}. In fact, Taiwan's semiconductor industry has witnessed double-digit growth for three consecutive years and is forecasted to reach more than 170 billion US dollars in 2022 \cite{itri2022}. The Taiwanese government heavily relies on the electronics industry for the next decades, as it plans on establishing several semiconductor initiatives and R\&D programs to support its growth \cite{itri2022}. In parallel, the Taiwanese government has issued a carbon reduction plan called the \textit{Greenhouse Gas Reduction Action Plan}, with the objective of decreasing their absolute carbon footprint at the 2050 horizon. Yet, Taiwan has been struggling with the effective implementation of this plan and actual carbon reductions are still to be realized~\cite{chou2019conundrums}. As the production of electronic components is known to be energy and resource-intensive~\cite{boyd2012life, bardon2020dtco, moreau2021could}, their environmental impacts cannot be ignored. This study hence addresses the question whether the fast-paced development of the electronics industry in Taiwan is consistent with the national sustainability roadmaps.

Assessing the sustainability of semiconductor manufacturing is not only important for Taiwan itself, but also for Western economies as both the USA and the European Union (EU) are currently massively investing to relocate the manufacturing of advanced manufacturing processes on their own soil~\cite{voas2021scarcity}. Indeed, the 2022 \textit{European Chips Act} projects more than 43 billion Euros of investments up to 2030 \cite{chips-act-2022}. It is clear that the EU aims at fostering a large digital innovation capacity based on a resilient semiconductor ecosystem. A key policy objective is to eventually bridge the gap from \textit{lab to fab} by producing CMOS chips in European foundries in order to secure its electronics supply chain and boost its competitiveness through digitalization. This is further supported by the recent chip shortage \cite{voas2021scarcity} which revealed the increasing pressure on our production systems. However, the Association for Computing Machinery (ACM) recently stressed the striking lack of environmental considerations in the EU Chips Act, the important risk of rebound effects and the absence of consideration for the carbon emissions linked to the unrestrained growth of ICT activities\cite{acm2022}. Although 195 countries have committed to strong carbon reduction targets by signing the 2016 Paris Agreement, the actual environmental impacts of electronic components over their whole life cycle is far from being thoroughly understood~\cite{pirson2022essderc}. It is therefore of critical interest to understand to which extent engaging into the relocalization of electronics manufacturing industry could impact the EU carbon reduction plans, especially if it focuses on the manufacturing of advanced semiconductor technologies. In this context, analyzing the current Taiwanese situation can provide a good estimate of the environmental impacts associated with the development of current and future state-of-the-art electronic components, given that more than 90\% of the worldwide sub-10nm integrated circuits (ICs) are currently manufactured in Taiwan at TSMC \cite{varas2021strengthening}, the \textit{de facto} world leader in advanced ICs manufacturing.

In this study, we estimate the environmental impacts of the electronics industry in Taiwan over the period 2015-2020 with a bottom-up approach. We use the CSR reports of 16 Taiwanese electronic components manufacturers (ECMs) to gather data over multiple environmental indicators such as final energy and electricity consumption, greenhouse gases (GHG) emissions, and water consumption. Finally, we discuss the conflicting trends between the Taiwanese sustainability roadmap and the rapid electronics industry development, through a scenario analysis. In addition, we use territorial perspectives to point out the terrain-related constraints preventing the island to implement a quick ecological transition.

The rest of the paper is structured as follows. Section~\ref{sec:background} presents similar studies and background while Section~\ref{sec:methodology} depicts the methodology behind this work. Results are then presented in Section~\ref{sec:results} for each environmental indicator and compared with national and industry data. Section~\ref{sec:discussion} discusses the link between increased production and environmental impacts, provides territorial perspectives, and details known limitations for this work. Finally, concluding remarks are given in Section~\ref{sec:conclusion}. 

\section{Background} \label{sec:background}

The manufacturing of electronic components is an energy-intensive industry that heavily uses electricity, chemicals and water~\cite{boyd2012life,bardon2020dtco}.
Electronic components are manufactured on high-purity silicon substrates using pure raw materials in clean-rooms, i.e., isolated spaces in which the air is continuously cleansed~\cite{krishnan2008hybrid}. Moreover, the fabrication processes often rely on perfluorocarbons (PFCs), gases that have a high global warming potential. These gases can however be abated (with abatement factors between 95--95\%~\cite{bardon2020dtco}) instead of being released in the atmosphere. Important quantities of electricity are needed to power the process tools, but also the facilities themselves (air ventilation and cleaning, water purification and cooling, ...)~\cite{hu2003power}. 
As an illustration, a life-cycle analysis studying the manufacturing of dynamic random access memories (DRAM) in Taiwan identifies global warming potential, non-renewable energy consumption but also the release of respiratory inorganics as major environmental impacts~\cite{liu2010life}. 

Over the years, Taiwan’s semiconductor industry has been widely studied for its industrial development based on the Science Park model, its economic efficiency and its R\&D policies. Hovewer, few works specifically look at the sustainability of this activity and the associated environmental impacts on the island.
A part of the literature analyzes how individual Taiwanese semiconductor companies try to improve the sustainability of their activities. 
Hu et al.~\cite{hu2016assessing} investigate corporate sustainability responsibility (CSR) reports of several semiconductor enterprises and link their decision making with the Sustainable Development Goals (SDG) of the United Nations.
Lin et al.~\cite{lin2018sustainability} propose a framework to evaluate the eco-efficiency and performance of a semiconductor company and apply it on several Taiwanese businesses. Hsu et al.~\cite{hsu2017investigating} survey Taiwanese semiconductor companies that are part of the Dow Jones Sustainability Index (DJSI) and identify their practices for engaging with a sustainability index. Such studies however do not provide an understanding of the overall environmental impacts of the semiconductor industry in Taiwan.

Among the rare studies that analyze nationwide the sustainability of the Taiwanese semiconductor industry, Chou et al. \cite{chou2019conundrums} provide the most up-to-date analysis regarding GHG emissions and electricity consumption.
The authors conduct their analysis using official statistics and past research by comparing the semiconductor industry with the petrochemical industry, which is the other largest industry on the island. Their findings indicate that, in 2016, these industries accounted together for 27.1\% of GHG emissions and 31.6\% of electricity consumption in Taiwan. The conclusion of the study remarkably underlines that \textit{"for the future of Taiwan’s economic development, the biggest challenge that the electronics industry face is whether it can bear the weight of its carbon emissions and electricity consumption"}~\cite{chou2019conundrums}.

On a broader perspective, Huang et al. \cite{huang2020planetary} compute absolute environmental sustainability indicators that assess how Taiwan performs with respect to several planetary boundaries (e.g., P\&N fertilizers, ocean acidification, freshwater use). This methodology is derived from the \textit{carrying capacity} concept, i.e., the maximum environmental interference a natural system can withstand without experiencing negative changes in structure or functioning that are impossible to revert~\cite{bjorn2016proposal}.
This study notably assesses the use of freshwater as high risk during the dry season in Southern and Central Taiwan, i.e., local natural boundaries regarding freshwater use are already exceeded in these parts of the island.

Incidentally, Chiu \cite{Chiu-2011} provides a comprehensive history of environmental movement against high-tech pollution in Taiwan, which largely includes the electronics industry. The author shows the important advances of the civil environmental movement from 2006 to 2011 but points out difficulties for this movement to move forward without government support.

Amongst the reviewed papers, the only publication that attempts to provide a national overview of the environmental impacts of the electronics industry is Chou et al.~ \cite{chou2019conundrums}.
Yet, the study only looks at GHG emissions and electricity consumption until 2016 using national aggregated data from government officials.
The authors also only provide a partial insight of the sustainability of the electronics industry by comparing it to the petrochemical one, and the critical issue of the water usage is not addressed.

To overcome these limitations, we gather in this work data directly from the CSR reports of several Taiwanese semiconductor companies over the 2015-2020 period. Our data set enables a more fine-grain analysis of the evolution of the GHG emissions, electricity consumption and water usage of several Tawainese ECMs. In addition, we use a territorial approach to account for geographical constraints regarding the stresses introduced on the production infrastructure.
This particular approach eventually allows us to conduct an in-depth discussion on the sustainability of Taiwan's electronics industry.
\section{Methodology} \label{sec:methodology}

In this section, we first describe which manufacturing activities are part of the scope of this study, and we then introduce the three studied environmental indicators. We finally explain how we generated a data set for the selected indicators using the CSR reports from 16 ECMs.

\subsection{Definition of the Study Scope}

\begin{figure}[h!]
    \centering
    \includegraphics[width=1\columnwidth]{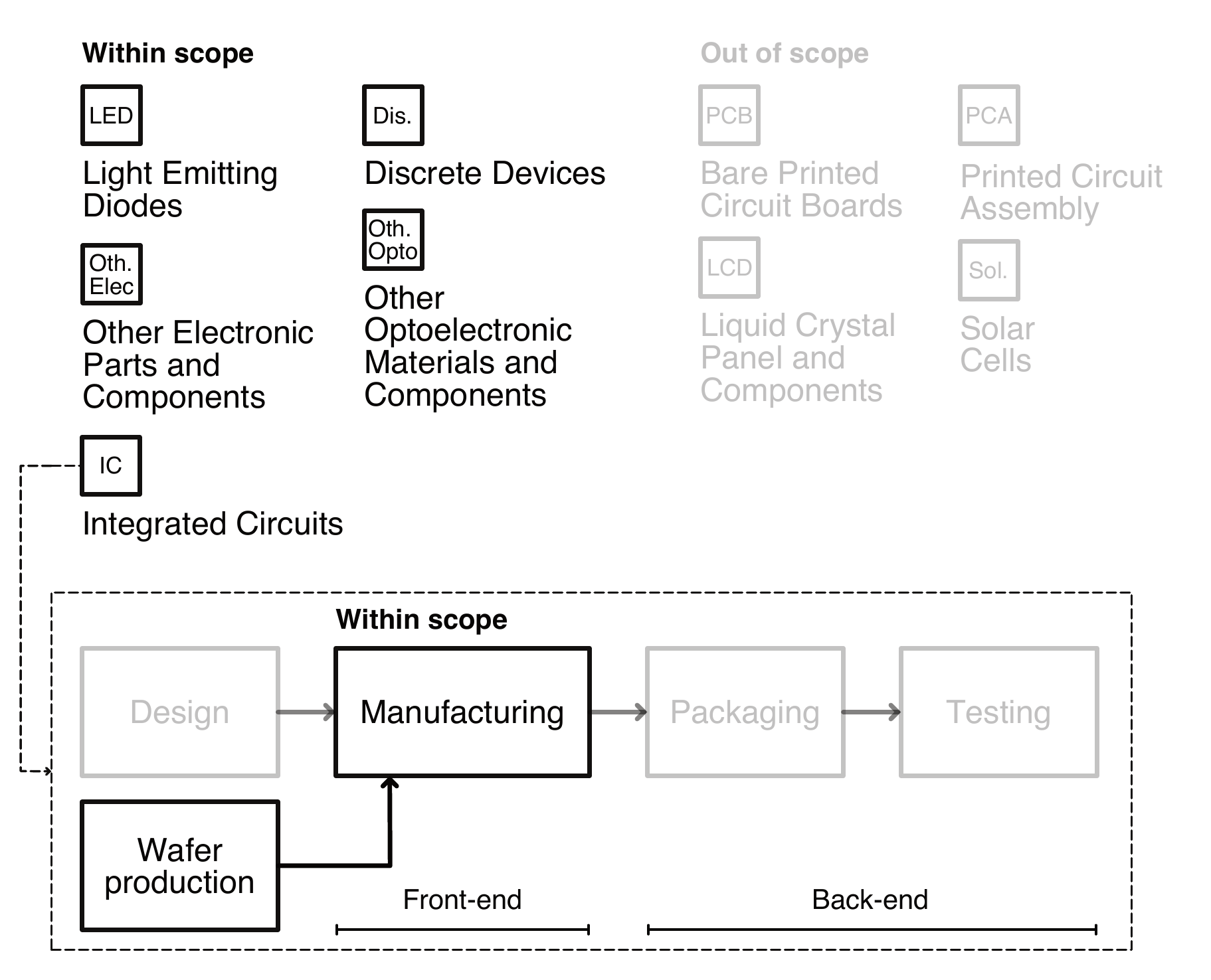}
    \caption{Simplified view of semiconductor manufacturing processes and the scope of our study (classification based on Taiwanese Indexes of production).}
    \label{fig:scope}
\end{figure}

Nowadays, ECMs manufacture a wide variety of electronic components with different technical specifications, which hence have different production processes. In this study, we restrict the analysis to the following categories of components, for which we were able to retrieve sufficient CSR data: integrated circuits (ICs), discrete devices, light emitting diodes (LEDs) and optoelectronic materials and components. Furthermore, these categories share similar manufacturing processes relying on the similar technologies, unlike printed circuit boards (PCBs) or liquid crystal panels (LCDs).


\begin{table}[h!]
\setlength\tabcolsep{3pt}
\caption{Listing of ECMs in Taiwan and selection of companies considered in this study.}
\vspace{0.2cm}
\renewcommand{\arraystretch}{0.9}
\centering
\label{tab:sample}
\resizebox{\columnwidth}{!}{%
\begin{tabular}{p{0.65\columnwidth} cccc}


\toprule[2pt]

 & \textbf{CSR} & \textbf{Taiwan} & \textbf{Within} & \textbf{Included} \\
\textbf{\large{Company name}} & data & territorial & scope & in this \\
 & & data &   & study \\

\toprule[2pt]

TSMC & \cgmark & \cgmark & \cgmark$^{(a)}$ & \cgmark \\
UMC & \cgmark & \cgmark & \cgmark$^{(a)}$ & \cgmark\\
Powerchip & \cgmark & \cgmark & \cgmark$^{(a,d,e)}$ & \cgmark\\
Vanguard & \cgmark & \cgmark & \cgmark$^{(a,b)}$ & \cgmark\\
Global Wafers & \cgmark & \cgmark & \cgmark$^{(f)}$ & \cgmark\\
Nanya & \cgmark & \cgmark & \cgmark$^{(a,e)}$ & \cgmark\\
Winbond & \cgmark & \cgmark & \cgmark$^{(a,e)}$ & \cgmark\\
Win Semiconductors & \cgmark & \cgmark & \cgmark$^{(a,e)}$ & \cgmark\\
Epistar & \cgmark & \cgmark & \cgmark$^{(c,d)}$ & \cgmark\\
Unimicron & \cgmark & \cgmark & \cgmark$^{(e)}$ & \cgmark\\
Innolux & \cgmark & \cgmark & \cgmark$^{(d)}$ & \cgmark\\
Nuvoton & \cgmark & \cgmark & \cgmark$^{(a,e)}$ & \cgmark\\
Lextar & \cgmark & \cgmark & \cgmark$^{(c,d)}$ & \cgmark \\
Everlight & \cgmark & \cgmark & \cgmark$^{(c,d)}$ & \cgmark \\
Optotech & \cgmark & \cgmark & \cgmark$^{(c,d)}$ & \cgmark\\
Wafer Works & \cgmark & \cgmark & \cgmark$^{(f)}$ & \cgmark\\

\midrule[0.2pt]

Siliconware Precision Industries & \cgmark & \cgmark & \xmark & \xmark \\
Foxsemicon Integrated Technology & \cgmark & \xmark$^\dagger$ & - & \xmark \\
Orient Semiconductor Electronics & \cgmark & \xmark$^\dagger$ & - & \xmark \\
Macronix & \cgmark & \xmark$^\dagger$ & - & \xmark \\
AUO & \cgmark & \xmark & – & \xmark \\
Global Foundries & \cgmark & \xmark & – & \xmark \\
Micron Memory Taiwan & \cgmark & \xmark & – & \xmark \\
Merck Performance Materials & \cgmark & \xmark & – & \xmark \\
Panjit & \cgmark & \xmark & – & \xmark \\
Lite-On Semiconductor & \cgmark & \xmark & – & \xmark \\
Delta Electronics & \cgmark & \xmark & – & \xmark \\
Advanced Semic. Engineering Inc. & \cgmark & \xmark & – & \xmark \\
Tatung Corp & \cgmark & \xmark & – & \xmark \\
Acer & \cgmark & \xmark & – & \xmark \\
Elan Microelectronics & \cgmark & \xmark & – & \xmark \\
Formosa Sumco Technology & \cgmark & \xmark & – & \xmark \\
Chipbond & \cgmark & \xmark & – & \xmark \\
Gudeng Precision & \xmark & – & – & \xmark \\
Episil & \xmark & – & – & \xmark \\
ProMOS & \xmark & – & – & \xmark \\
Taiwan Semi & \xmark & – & – & \xmark \\
Arima Optoelectronics & \xmark & – & – & \xmark \\
AWSC & \xmark & – & – & \xmark \\
SAN CHIH Semiconductor & \xmark & – & – & \xmark \\
Creative Sensor Inc. & \xmark & – & – & \xmark \\
Visera Technologies (Part of TSMC) & \xmark & – & – & \xmark \\
Episil & \xmark & – & – & \xmark \\
Creating Nano & \xmark & – & – & \xmark \\
Episil-Precision & \xmark & – & – & \xmark \\
Sun Yuan Technology & \xmark & – & – & \xmark \\
U-Can Dynatex & \xmark & – & – & \xmark \\
Ofuna Technology & \xmark & – & – & \xmark \\
Cin Phown Technology & \xmark & – & – & \xmark \\
Davicom Semiconductor & \xmark & – & – & \xmark \\
Holtek Semiconductor & \xmark & – & – & \xmark \\
Realtek Semiconductor & \xmark & – & – & \xmark \\
WiseChip Semiconductor & \xmark & – & – & \xmark \\
Unikorn Semi & \xmark & – & – & \xmark \\
Sigurd Microelectronics & \xmark & – & – & \xmark \\
Mosel Vitelic & \xmark & – & – & \xmark \\
Kneron & $\ominus$ & – & – & \xmark \\
Green Energy Technology & \xmark & – & – & \xmark \\
Sino-American Silicon Products & \xmark & – & – & \xmark \\
Hexawave & \xmark & – & – & \xmark \\

\midrule[0.2pt]

\textbf{COUNT:} \hspace{3cm} 60 & 33 & 20 & 19 & \textbf{16} \\

\bottomrule[2pt] \\

\multicolumn{4}{l}{ \cgmark : Yes \quad \xmark : No \quad $\ominus$ : not relevant \quad $\dagger$: not enough data \quad - : not checked
} \\

\multicolumn{4}{l}{(a) : IC \quad (b) : Disc. \quad (c) : LED \quad (d) : Oth. Opto. \quad (e) : Oth. Elec. \quad (f) : Wafer
}

\end{tabular}%
}

\end{table}

The manufacturing of the electronic components under study implies complex supply chains with several actors located in different countries. For instance, Figure~\ref{fig:scope} illustrates a simplified view of CMOS ICs manufacturing. The chain could be divided in two main parts, the front-end and the back-end processing. The front-end processing corresponds to the fabrication of the electronic components with a semiconductor process. Such process usually involves several hundreds of steps that modify the wafer to give it the desired electrical properties. In the back-end, the many circuits contained in a single processed wafer are cut into silicon dies and packaged in independent chips. The packaged chips are then finally tested. These stages usually take place at different geographical locations, involving transportation and operations by several companies. As a first step to map this complex ecosystem, we narrow down the scope of this study by excluding back-end processing which is generally less impacting than front-end manufacturing, as suggested by \cite{prakash2013schaffung}. We further narrow the scope of this study by excluding design and raw material extraction as we focus exclusively on the environmental impacts of ECMs that are geographically located on the island of Taiwan. Note that for optoelectronics companies, the distinction between front-end and back-end processing is not always clear, as pointed out in  Table~\ref{tab:sample}. Yet, they are included in this study as they represent a non-negligible part of the environmental impacts (as shown in the supplementary material).

\subsection{Selection of the Environmental Indicators}

We consider in this study three indicators that quantify the main environmental impacts of ECMs identified in Section~\ref{sec:background}: GHG emissions, final energy consumption, and water usage. These indicators are nowadays often included in CSR reports. The GHG emissions of an ECM are quantified in kg\coo -equivalent (kg\coo eq). CSR reports follow the GHG Protocol Corporate Standard to report their GHG emissions~\cite{ranganathan2004greenhouse}. This protocol classifies the emissions of a company into three scopes. 
Scope 1 emissions are direct emissions from owned or controlled sources, while Scope 2 emissions are indirect emissions from the generation of purchased energy (mainly electricity). Scope 3 accounts for all the indirect emissions outside of Scope 2 that occur in the value chain of the company, e.g., the emissions released by the refining of raw materials by a supplier. Since we conduct a territorial analysis that only comprises the environmental impacts on island of Taiwan, we exclude Scope 3 emissions from our GHG indicator. On the contrary, since Taiwan is an island with an independent electricity grid (i.e., all the electricity is generated on the island), Scope 2 emissions are taken into account.

We will see in Section~\ref{sec:results} that most of the GHG emissions of Taiwanese ECMs are actually indirect emissions from the electricity production. It is therefore important to quantify the global energy usage (in TWh) of the industry, and in particular the electricity consumption. Our indicator combines both the final energy generated from fossil and renewable sources, as well as the energy obtained from electricity purchases.
We notably discuss in Section~\ref{sec:discussion} the proportional role of each energy source in the electronic components manufacturing in Taiwan. 

Finally, our last indicator evaluates the amount of water in cubic meters (m³) consumed by a company. Let us mention that the difference between water use and net water consumption is not always clear in CSR reports, partly due to the different recovery and discharge water processes that take place in semiconductor facilities. Reports rarely explicitly detail whether water is taken from the tap, recycled, reused or reclaimed, but the reported data still provides a good insight on the water consumption of ECMs.

Nevertheless, even if the three aforementioned indicators already capture many aspects of the environmental impacts of semiconductor manufacturing, they fall short of providing an exhaustive picture of all environmental interference. For instance, toxicity concerns, biodiversity loss and impacts on ecosystems are not investigated in this study, as well as air pollution and wastewater, even though they represent a historical environmental problem on the island \cite{Chiu-2011}.
Further research on these issues is hence still needed.

\subsection{Data Collection}

The list of ECMs selected in this study has been obtained by cross-referencing databases from two industry associations, namely the Taiwan Semiconductor Industry Association\cite{TSIA-2021} and the SEMI Global industry association for electronics manufacturing and design supply chain\cite{SEMI-2021}. Each database discloses a list of ECMs with their specific industrial activities.
The research criteria used to identify ECMs belonging to our study scope were: "Manufacturing", "Semiconductor", "Semiconductor, "Photonics" as activity and "Taiwan" as location. A total of 60 companies fulfilled these criteria, but only 30 of them had published environmental data in CSR reports and 20 had disclosed location-based data specific to Taiwan. In the end, our sample contains 16 companies with workable data, as shown in Table~\ref{tab:sample}. The data set gathers mainly companies focusing on ICs, LEDs and optoelectronic components manufacturing.

After the selection of the relevant Taiwanese ECMs, we created a data set by compiling their environmental impacts using the annual CSR reports from each company.  Thanks to the Global Reporting Index standards\cite{gri-home}, compliant CSR reports now contain company-specific data about the three selected environmental indicators (GHG emissions, final energy and electricity consumption and water usage). For each company in our sample, we retrieved the values for the period 2015-2020 of the three indicators and compiled them in a database. This database is provided as Supplementary Materials.
We selected the 2015-2020 period because environmental data reporting became more widespread for Taiwan's industry since 2015, and because CSR reports for the year 2021 were not yet available at the time of writing.

Finally, we conduct in Sections~\ref{sec:results} and~\ref{sec:discussion} comparisons with the aggregated data from our sample with public data at the industrial and national levels provided by the Taiwanese government and by the national energy provider, Taipower.

\section{Results} \label{sec:results}  

This section presents the aggregated data obtained from our sample of 16 Taiwanese ECMs, with respect to the three environmental indicators defined in Section~\ref{sec:methodology}.
For each indicator, the compound annual growth rate (CAGR) is given to evaluate its annual rate of increase. A CAGR of 15\% over the period under study (i.e., 5 years) corresponds to a doubling of the corresponding environmental impact.
The footprints of the electronics sub-sector are then compared to those of other Taiwanese sectors.

It is important to point out that, in the Taiwanese ecosystem, the company TSMC is by far the largest semiconductor manufacturer, with a revenue in 2020 of NT\$ 1,339 billions. As a comparison, the second largest EMC, UMC, only attained a revenue of NT\$ 176 billions during the same year~\cite{tsia-overview}. Since the production of TSMC dwarfs those of other manufacturers, the contribution of TSMC to the aggregated indicators is shown separately from the rest of the sample to provide a more fine-grain analysis of the overall environmental trends.

\subsection{GHG Emissions}

\begin{figure}[t]
    \centering
    \includegraphics[width=1\columnwidth]{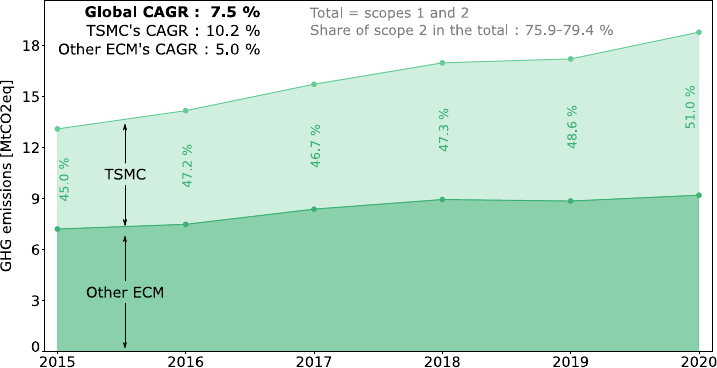}
     \caption{GHG emissions (Scopes 1 and 2) over the period 2015-2020 for the ECMs considered in this study. The distinction is made between TSMC and the other ECMs (the percentages indicate the share of TSMC's emissions).}
    \label{fig:absolute-GHG}
\end{figure}

We first present the evolution of the GHG emissions from the ECMs in our sample. Fig.~\ref{fig:absolute-GHG} shows the GHG emissions belonging to the Scopes 1 and 2, i.e., the direct emissions and the indirect emissions due to purchased energy, respectively.
The data shows a strong 43.3\% increase of GHG emissions during the period 2015-2020 (+43.3\%), which corresponds to a CAGR of +7.5\%. Although TSMC significantly contributes to the overall trend (CAGR of 10.2\%), the share of GHG emitted by the other ECMs in the sample also clearly increased, with a CAGR of +5.0\%.

Among the overall GHG emissions, the indirect emissions linked to energy purchases (i.e., Scope 2) largely dominate, with shares that vary between 75.9 and 79.4\% over the years. We show next that the Scope 2 emissions originate from electricity consumption, which in Taiwan is mainly generated from highly carbon-intensive sources.

The remaining 20.6-24.1\% of Scope~1 emissions are hence direct emissions released from the EMCs facilities,  
including on-site electricity generation and emissions of high GWP gases such as fluorinated compounds (e.g., SF$_6$, NF$_3$, CF$_4$, CHF$_3$), that are used for processing or maintenance steps in the manufacturing processes~\cite{bardon2020dtco}.
Yet, very interestingly, it can be observed in the data that for most ECMs, Scope~2 emissions increase much faster than those of Scope~1.
Scope 2 emissions of TSMC increased by 72\% over the studied period, whereas its Scope 1 emissions grew only by 37\%. For UMC, Scope~1 emissions even decreased by 13\% while Scope~2 emissions increased by 10\%.
These trends indicate that Taiwanese ECMs tend to invest in efficient abatement systems to mitigate their direct emissions of fluorinated compounds. The latest and most efficient abatement systems reach abatement factors near 99\%, whereas previous generations usually reached a 95\% efficiency~\cite{bardon2020dtco}.
From these observations, it seems that a further decoupling of the volume of manufactured ICs from their direct GHG emissions will be very difficult and costly to achieve.

\subsection{Energy and Electricity Consumption}

\begin{figure}[t]
    \centering
    \includegraphics[width=1\columnwidth]{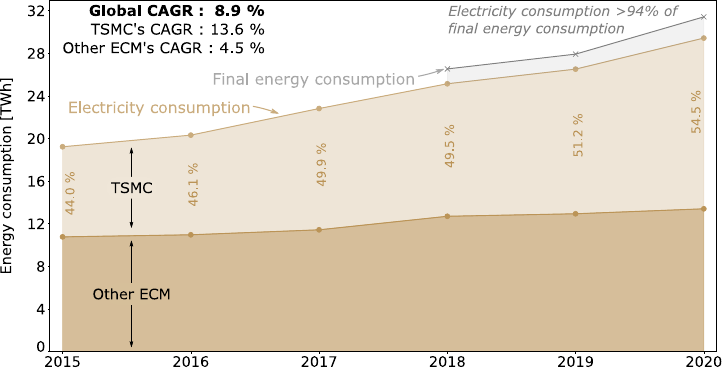}
     \caption{
     Energy and electricity consumption over the period 2015-2020 for the ECMs considered in this study. The distinction is made between TSMC and the other ECMs (the percentages indicate the share of TSMC's emissions). CAGRs are given for the electricity consumption.}
    \label{fig:absolute-energy}
\end{figure}

Since indirect emissions due to energy purchases dominate the overall GHG emissions of Taiwanese ECMs, it is also crucial to study the evolution of their energy consumption.
Fig.~\ref{fig:absolute-energy} shows that the final energy consumption of the 16 ECMs in our sample has increased by +18.4\% over 3 years\footnote{Five ECMs did not report their final energy consumption from 2015 to 2017, but well their electricity consumption. We hence show the total final energy consumption only for the period 2018-2020}, corresponding to a CAGR of +8.8\%. In particular, TSMC accounts for more than half of the energy consumption among the sampled ECMs and has increased its final energy consumption by 90\% over the last five years. More importantly, we observe that electricity is the main source of energy for ECMs, with shares greater than 94\% of the global final energy use.  
As shown in Fig.~\ref{fig:absolute-energy}, the total electricity consumption has increased by +53.3\% over the studied period, yielding a CAGR of +8.9\%. 
This trend is observed for all but two ECMs. Only Lextar and Nuvoton have slightly reduced their electricity consumption between 2015 and 2020, respectively with -29\% and -8\%. Unfortunately, these improvements have no effect on the general trend as they are achieved by non-dominant actors. 

The joint analysis of both Fig.~\ref{fig:absolute-GHG} and Fig.~\ref{fig:absolute-energy} indicates that the GHG emissions of the studied ECMs are strongly correlated to their electricity consumption~\cite{chou2019conundrums}.
This correlation stems from the fact that electricity in Taiwan is mainly generated from highly carbon-intensive energy sources such as coal (45.3\%), natural gas (36\%) and oil (1.5\%) while nuclear (11.2\%) and renewable energies have much smaller contributions (5.9\%) \cite{IEA-Taipei}.
An abundant access to zero-carbon electricity will therefore be needed to reduce the indirect GHG emissions of Taiwanese ECMs. This issue is discussed in more details in Section~\ref{sec:discussion}.

\subsection{Global Water Usage}

We now present the aggregated water usage for the ECMs belonging to our data set. We report in this section global values corresponding to the entire island of Taiwan. However, contrary to the previous indicators, water-related issues should rather be considered with a territorial approach. Since water is much more difficult to transport than electricity, water shortages are often restricted to specific regions and a global indicator does not convey this reality. We hence also conduct in Section~\ref{sec:discussion} a territorial analysis of the water usage on the island of Taiwan, and its consequences for the semiconductor industry.

\begin{figure}[t]
    \centering
    \includegraphics[width=1\columnwidth]{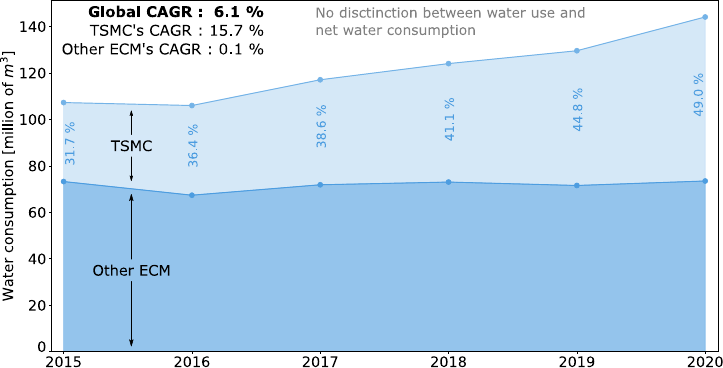}
     \caption{
     Water usage over the period 2015-2020 for the ECMs considered in this study. The distinction is made between TSMC and the other ECMs (the percentages indicate the share of TSMC's emissions).}
    \label{fig:absolute-water}
\end{figure}

Figure~\ref{fig:absolute-water} shows the amount of water consumed by the sampled ECMs. The global volume increased by +34.4\% over 6 years, corresponding to a CAGR of +6.1\%. This global increase comes exclusively from TSMC, which used 70.6 million m$^3$ of water in 2020 and increased its consumption by 108\% over 6 years.
This highlights that even though three EMCs have reduced their water consumption between 2015 and 2020, namely Innolux with -16\%, Unimicron with -20\%, and Winbond with -48\%, these savings have almost no effect on the global indicator because of TSMC's considerable increased water usage.

\subsection{National and Industry-Wise Comparison}
\label{subsec:comparison}

We finally analyze the aggregated environmental footprint of the 16 ECMs we studied by comparing it to the footprints of other Taiwanese sectors (industry, residential, services, transport, agriculture and energy) for the year 2020. 
This systematic comparison allows a better understanding of the importance of the environmental impacts of ECMs both at the industrial and national levels.

\begin{figure*}[t]
    \centering
    \includegraphics[width=1\textwidth]{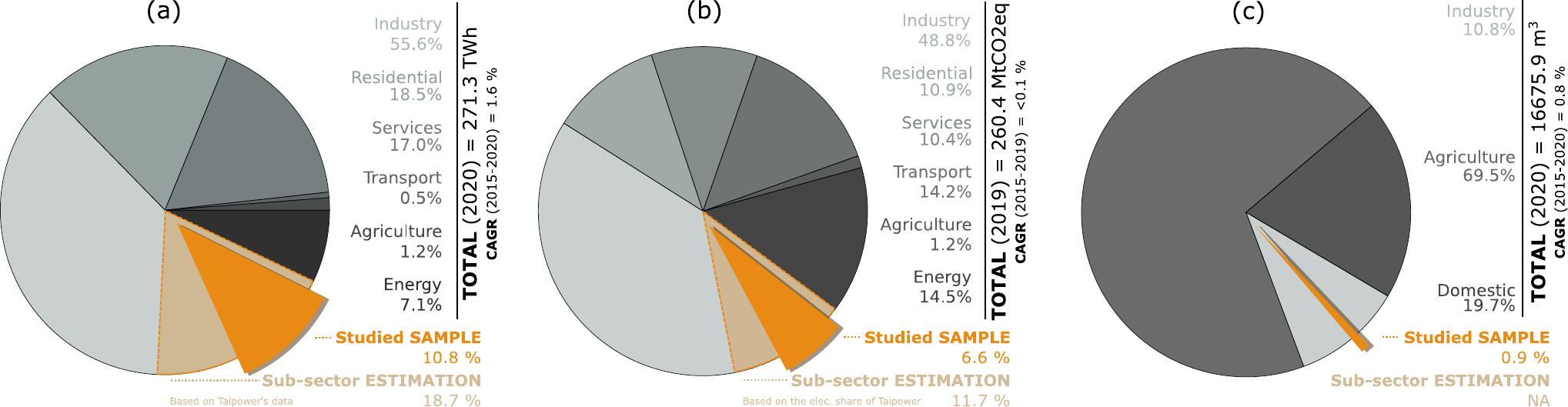}
     \caption{National shares of (a) electricity consumption, (b) GHG emissions, and (c) water consumption in 2020 for the different Taiwanese sectors. The contribution of the ECMs from our sample belong to the electronics sub-sector, which itself is part of to the industrial sector.}
    \label{fig:shares}
\end{figure*}

We first discuss the weight of electronics manufacturing in the overall Taiwanese electricity consumption.
The Taiwanese Ministry of Economic Affairs reports annually the quantity of electricity consumed by each sector~\cite{BoE-2021}. Fig.~\ref{fig:shares} shows the proportions of each sector in the Taiwanese electricity consumption. In addition, Taipower, the dominating electricity supplier for the electronics sub-sector, annually reports the electricity consumption of this sub-sector under a category named \textit{Electronic Components Manufacturers}~\cite{NDC-2021}. The industry represents 55.6\% of the overall electricity consumption and, according to Taipower, its electronics sub-sector uses a striking 18.7\% of the same national consumption. Our sample of 16 ECMs accounts for 10.8\% of the national electricity budget, i.e., approximately a fifth of the entire sector. The important weight of our sample in the sub-sector (56 to 70\%) indicates that, despite studying only 16 out of the 60 Taiwanese ECMs, our set of selected companies still represents a representative part of the targeted sub-sector thanks to major actors such as TSMC or UMC.

Regarding the emissions of GHG, the industrial sector is responsible for 48.8\% of the 260.4 Mt of \coo eq released in Taiwan during 2020.\cite{BoE-2021} In this sector, our sample of 16 ECMs represents 6.6\% of the annual emissions. 
Since there are no official statistics for the GHG emissions specific to the electronics sub-sector, we propose the following method to extrapolate the carbon footprint of the sub-sector from our sample. We previously observed that the studied ECMs have GHG emissions strongly correlated to their electricity consumption, since the Scope~2 of these companies outweights their Scope~1 emissions~\cite{TSMC-2020, UMC-2020, Innolux-2020}.
By assuming that this correlation holds for all Taiwanese ECMs (they all rely on the same carbon-intensive electricity sources), we estimate the GHG emissions of the entire electronics sub-sector by scaling the aggregated GHG emissions from our sample, which represents 58\% of the electricity consumed in the sub-sector, to the overall electricity usage disclosed by Taipower. This extrapolation yields an estimated annual share of 11.7\% of the national GHG emissions.
This result confirms the important responsibility of the Taiwanese ECMs in the national GHG emissions, which has also been raised in~\cite{chou2019conundrums}. The limitations of the proposed extrapolation method are discussed in Section~\ref{subsec:limits}.

Finally, Fig.~\ref{fig:shares} also shows the proportional share of the studied ECMs in the national water usage. We have no straightforward method to estimate the water usage of the entire sub-sector for this indicator. The 16 ECMs surprisingly consume only 0.9\% of the national water usage, while the agricultural sector uses up to 70\% of the total.
Yet, as previously discussed, even if the national footprint of ECMs is only a small part of the total water consumption, their water usage is highly concentrated in a few Science Parks. We hence conduct a local territorial analysis in the next section to better assess the risks of water shortages for the electronics sub-sector.
 
\section{Discussion and Territorial Analysis} \label{sec:discussion}

In this section, we first compare the absolute increase of the environmental indicators highlighted in Section~\ref{sec:results} with national statistics of the production volume.
We subsequently leverage the results of our study to question the fast-paced development of the Taiwanese semiconductor manufacturing sector and its consequences for the national climate roadmap.
Finally, we discuss the limits of this study and we point out some blind spots in our analysis. 

\begin{figure}[t]
    \centering
    \includegraphics[width=1\columnwidth]{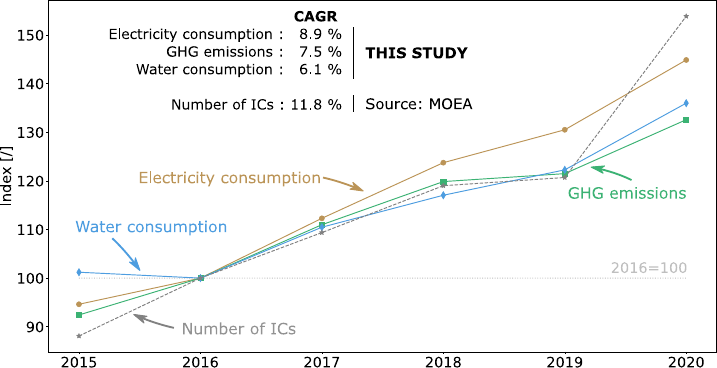}
    \caption{Evolution of the environmental indicators obtained in this study and of the ICs manufactured in Taiwan over the period 2015-2020. All data is normalized with respect to 2016, i.e., 2016$=$100.}
    \label{fig:index}
\end{figure}

\begin{figure*}[h!]
    \centering
    \includegraphics[width=0.92\textwidth]{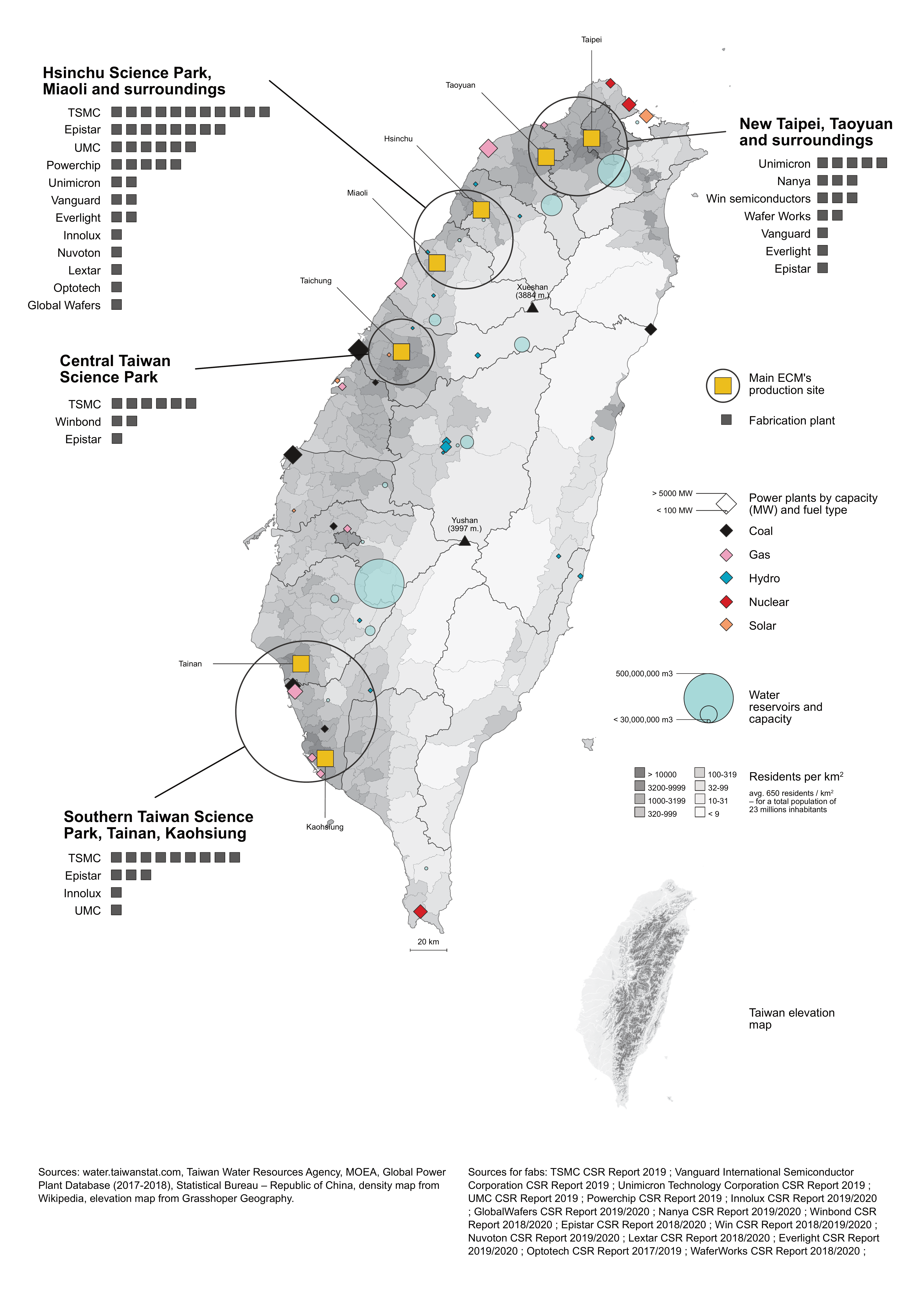}
     \caption{Location of companies considered in this study together with power plants and water reservoirs compared to population density.}
    \label{fig:map}
\end{figure*}

\subsection{Correlations Between Environmental Indicators and Production Volume} \label{sec:hypo}

We observed in Section~\ref{sec:results} that all three environmental indicators of 16 Taiwanese ECMs have significantly increased over the studied period. Possible explanations for these trends could reside either in an increase in the production volumes or in an increase of the resource intensity per component over the years. To obtain a better insight over these dynamics, Fig.~\ref{fig:index} depicts the normalized evolution of the three aggregated indicators and of the number of manufactured ICs in Taiwan from 2015 to 2020, according to data from the Ministry of Economic Affairs \cite{MOEA-2021}.
We focus specifically on the manufacturing of ICs because it was not possible to match perfectly the scope of our study with the available data from the Ministry of Economic Affairs. Nevertheless, Table~\ref{tab:sample} points out that a significant share of companies under study are specialized in IC manufacturing.
We observe that the number of ICs produced in Taiwan follows an upward trend which is very similar to the trend of our indicators. Albeit further work is needed to properly identify the main reasons behind the rise of the environmental indicators of Taiwanese ECMs, all three indicators are clearly correlated to the national production volume.

These correlations raise questions regarding the inherent sustainability of the Taiwanese electronics industry. Indeed, they suggest the impossibility of an absolute decoupling between production volume and environmental impacts, and in particular, of GHG emissions~\cite{bol2021moore}.
In addition, the development of advanced CMOS sub-10nm technologies could further increase the carbon intensity per component~\cite{pirson2022essderc}. As an example, more advanced manufacturing technologies such as extreme ultraviolet lithography (EUV) are clearly expected to increase electricity consumption by approximately +50\% compared to current mainstream manufacturing processes~\cite{chou2019conundrums}.
We thus discuss next whether Taiwan could sustain the development of its semiconductor industry while complying with their national carbon reduction plans.

\subsection{Territorial Constraints of Taiwan for Electricity and Water Supply}

To further study the long-term sustainability of the Taiwanese electronics industry, we now analyze the territorial constraints of Taiwan regarding renewable electricity generation and water supply.
The following discussion should hence be considered with Taiwan's geography in mind. Taiwan is a small country (36,197 km$^2$) with a mostly mountainous terrain on the eastern side of the island, as shown in Fig.~\ref{fig:map}. Because of this particular geography, most of the population is gathered on flat regions near the west coast. Even though the average population density is about 650 residents/km$^2$ for a total of 23 million people in Taiwan \cite{pop-density-taiwan}, some urban regions of the island attain a population density greater than 10~000 residents/km$^2$. In additional, urban areas, energy and water sources and industries tends to be concentrated in the same areas. The Taiwanese topography suggests that the future development of renewable energies, as future industry facilities, is ultimately constrained by the remaining available space on the island. 

We observed in Section~\ref{sec:results} that indirect emissions due to electricity consumption are the major source of global warming contribution from ECMs. Indeed, up to 97.5\% of Taiwan's energy generation originates from imported carbon-intensive energy sources in 2020, including coal from Australia, Indonesia and Russia, crude oil from Saudi Arabia, Kuwait and US and liquefied natural gas (LNG) from Qatar, Australia and Russia \cite{BoE-2021}. Regarding specifically the production of electricity, the Taiwanese government plans to phase out nuclear power and to produce 20\% of its electricity from renewable sources by 2025~\cite{gao2021review}, under the so-called \textit{2017 Electricity Act}. Yet, renewable energy represented less than 6\% of the power generation in 2020~\cite{tsai2021trend, IEA-Taipei}. In fact, as of 2020, only 9.47~GW of renewable energy capacities have been installed in Taiwan (including 5.82~GW from photovoltaic power and 0.85~GW from onshore and offshore wind power)~\cite{tsai2021trend}. Renewable energies generated that year 15.12~TWh of electricity, whereas the sole electronics sub-sector consumed 50.73~TWh. Due to the limited usable space on the island, the Taiwanese government mainly counts on offshore wind turbines in the future, with a target capacity of 5.7~GW by 2025. Yet, by June 2021, only 0.13~GW have been installed~\cite{gao2021review}. In addition, controversies related to the establishment of a full-scale offshore wind project have continued to occur since mid-2017, undermining the shift towards renewable energy sources in the short term. It becomes hence extremely challenging for Taiwain to reach its 2025 targets and ultimately to decarbonize its electricity production~\cite{tsai2021trend, chou2019conundrums}.

Analyzing Taiwan's geography is also crucial to understand the water management and the water storage infrastructure present on the island. Even if ECMs do not represent a significant part of the national water intake, they rely on a limited number of reservoirs located in places of very high population density as depicted in Fig.~\ref{fig:map}. For instance, TSMC's factories in Hsinchu Science Park use 10.3\% of the daily supply from Baoshan and Second Baoshan reservoirs whereas those located at STCP (Southern Taiwan Science Park) use 5.3\% of the daily supply from Nanhua and Zengwen reservoirs \cite{TSMC-2020}. This concentration has already led to supply issues during droughts. In 2020, the island faced a drought caused by the absence of typhoons in the previous year. If ECMs were already facing stress in the water-intensive production chain, this drought seriously amplified the crisis in this key sector of the island. Indeed, TSMC relied that year on hundreds of water trucks to maintain its water supply while several restrictions where being taken at the national level\cite{reuters}. In this sense, Taiwanese ECMs share a liability regarding water supply due to the increasing and spatially concentrated demand in the Science Parks.

\subsection{Perspectives With Respect To Taiwan's Climate Roadmap}

\begin{figure}[t]
    \centering
    \includegraphics[width=1\columnwidth]{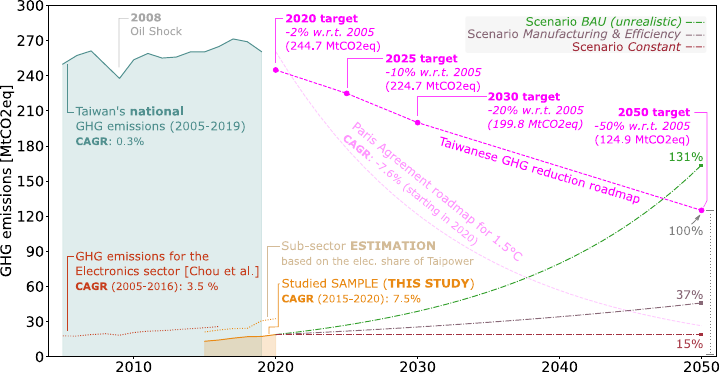}
    \caption{Scenario analysis for ECMs' GHG emissions in Taiwan from 2020 to 2050. Perspectives regarding the national GHG emissions and the Taiwanese carbon reduction roadmap. The share of each scenario with respect to the total carbon budget in 2050 is given on the bottom right.}
    \label{fig:pathways}
\end{figure}

While Taiwan's GHG emissions ranged from 238 to 271 MtCO2e between 2005 and 2019, the country plans to decrease GHG emissions compared to 2005 by 20, 30, and 50\% in 2025, 2030, and 2050, respectively. Consequently, the total GHG emissions 2050 target for Taiwan is about 125 MtCO2e, as illustrated in Fig.~\ref{fig:pathways}. This could be seen as the available Taiwanese carbon budget in 2050. Yet, the Taiwanese roadmap is based on linear reductions over 30 years, which is significantly least demanding than the exponential reduction in the roadmap that has been negotiated in the Paris Agreement for the 1.5$^\circ$C target, i.e., a reduction of -7.6\%/year if started in 2020 \cite{unepGAP-2019}. \par

To contribute to the national target, Taiwanese ECMs will have to mitigate their GHG emissions. We study three scenarios depicted in Fig.~\ref{fig:pathways} to assess the coherency of the GHG evolution from the electronics sub-sector with the Taiwanese climate roadmap. In general, projections for the environmental footprint of electronics rarely look ahead by more than 15 years mainly because of the high uncertainty regarding the evolution of technology and uses \cite{freitag2021real, Pirson2021}. Nevertheless, the Taiwanese roadmap draws a GHG reduction roadmap up to 2050 which motivates a long-term projection in this case. Here, we only use conservative assumptions to design these scenarios. We do not assume improvements or degradation of the efficiency of manufacturing processes, since such hypotheses strongly depend on the type of products that are considered. In addition, the scenarios use the 2020 values from our sample of 16 ECMs, and not the extrapolated value for the whole sub-sector.

The first scenario aims at illustrating the GHG emissions that would result froma \textit{business as usual} (BAU) policy. It assumes that the GHG emissions of EMCs will increase at the same pace than observed for the sample under study over the period 2015-2020, i.e., with a CAGR of 7.5\%. This scenario illustrates that maintaining the current pace of development is incompatible with the national climate targets. The carbon footprint of the ECMs sample would indeed surpass the total carbon budget of Taiwan in 2050. Furthermore, it is extremely unlikely that Taiwan could continue to produce an ever-growing number of components at an exponential pace when considering practical constraints, especially when considering the energy consumption.

The second scenario called \textit{manufacturing \& efficiency} assumes the evolution of GHG emissions as depicted by Chou et al. \cite{chou2019conundrums} for the electronics sector over the period (2005-2016), i.e., 3.5\%. In addition, this scenario also assumes best-effort improvements by Taiwanese electricity suppliers such that the carbon intensity of electricity decreases by -0.6\%/year, as suggested in \cite{chou2019conundrums}. We apply these improvements only on Scope 2 as they are associated with electricity consumption. This scenario illustrates that the carbon footprint of the sample would reach up to 37\% of the available carbon budget, which is significantly higher than today.

Finally, the last scenario considers that the ECMs keep their absolute carbon footprint unchanged from 2020 onward, independently of the evolution of the production volume and the electricity carbon intensity. In this \textit{constant} scenario, the ECMs in our sample would represent about 15\% of the total carbon budget in 2050.

The results in Fig.~\ref{fig:pathways} point out the inconsistency between the GHG emissions from the electronics sub-sector and Taiwan's GHG roadmap. Indeed, following the different scenarios, at least 15\% to 37\% of the 2050 carbon budget would be allocated to the manufacturing of electric components. These percentages should even be revised upwards when considering the entire sub-sector. In all cases, our analysis clearly highlights the difficult upcoming political arbitrations that will be needed if ECMs cannot decrease their GHG emissions.

\subsection{Towards a Potential Carbon Lock-in}

The previous discussions suggest that there is a very low likelihood for Taiwanese ECMs to fully decarbonize their manufacturing in a foreseeable future.
We finally argue that this sub-sector is likely to suffer from a \textit{carbon lock-in}.
A carbon lock-in describes a situation where, in a economical context of returns to scale, both public institutions and private actors (which constitute, with the deployed technological infrastructure, a \textit{techno-institutional complex}) drastically inhibit the competitiveness and roll-out of low-carbon alternatives~\cite{unruh2000understanding}.

\begin{figure}[t]
    \centering
    \includegraphics[width=1\columnwidth]{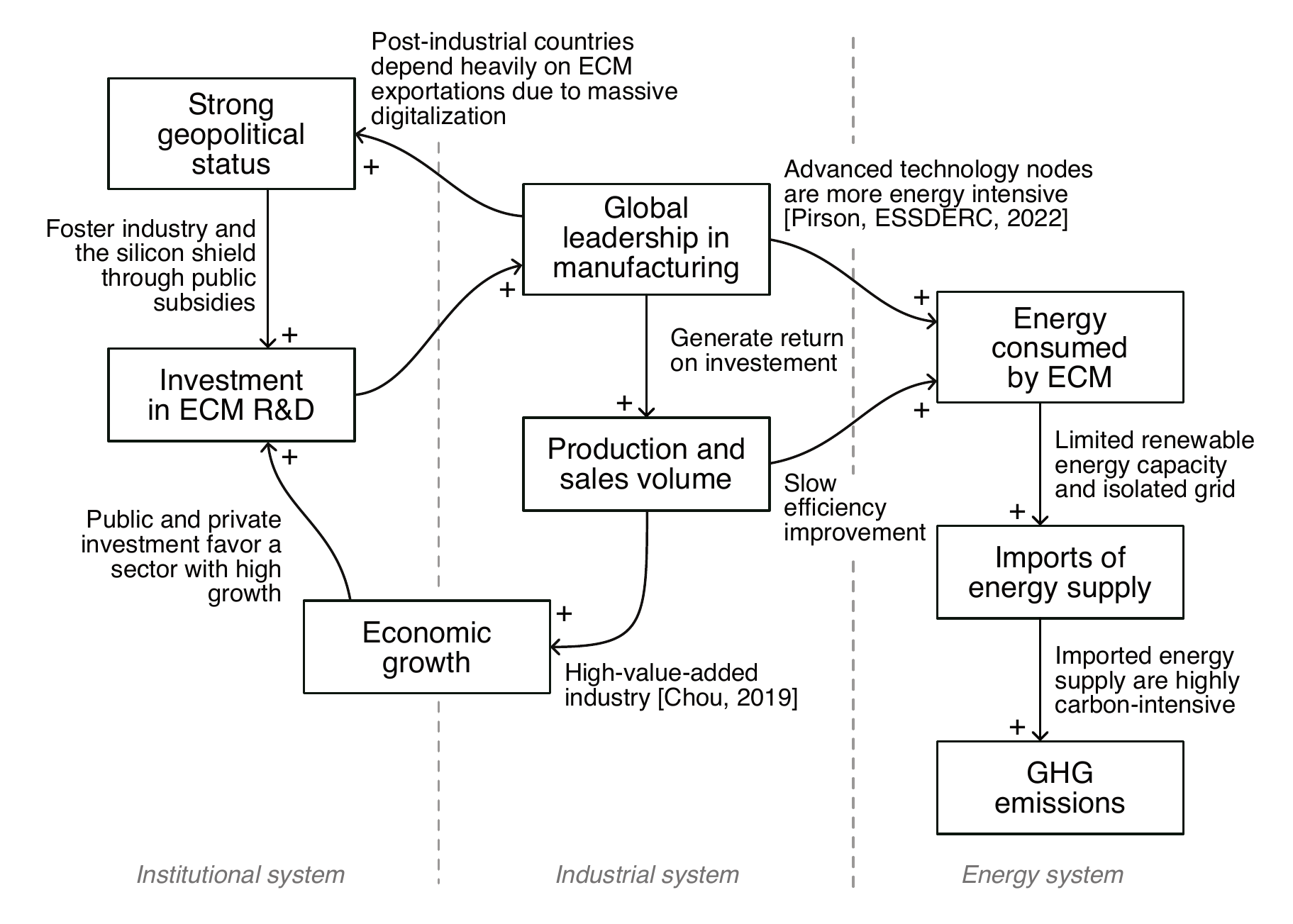}
    \caption{Illustration of the techno-institutional complex fostering a lock-in on carbon-intensive energy sources.}
    \label{fig:lock}
\end{figure}

We first point out that the Taiwanese geopolitics and economy are mostly centered around its electronics industry. The Taiwanese \textit{silicon shield} is a well-described geopolitical strategy against the ambitions of mainland China on the island. This strategy relies on being a worldwide leading supplier of electronic components, enabling to seek the military protection of the United States~\cite{addison2001silicon}. To reach this crucial leading position, the Taiwanese government fostered the electronics industry by creating the original model of Science Parks, and keeps to heavily subsidizing it~\cite{MOEA-incentive}. In return, leading Taiwanese ECMs have also committed substantial investments, e.g., TSMC is investing up to 44 billion USD in 2022 to increase production and R\&D efforts~\cite{WSJ}. Yet, the technological path towards advanced sub-10nm technologies also implies new industrial processes such as EUV, which in turn are likely to increase the environmental footprint and energy consumption of the industry (the energy consumption from EUV processes is more than ten times that of deep ultraviolet (DUV) processes~\cite{TSMC-2020}). At the same time, the territorial analysis previously discussed showed the difficulties for Taiwan to conduct a large-scale deployment of renewable energies in the short and medium-term. We therefore expect that a further grow of the production volume of ECMs will require increasing imports of high-carbon energy supplies. When considering these interactions together, as shown in Fig.~\ref{fig:lock}, it becomes apparent that the Taiwanese electronics industry is fated to remain a major contributor of GHG emissions.

From this analysis specific to Taiwan, lessons can be learned for other countries that want to develop a sub-10nm industry, such as the European Union through the Chips Act or the USA. While Taiwan suffers from carbon-intensive energy sources and a constrained territory, the European Union could place its future industry in places where power is less carbon-intensive and can also disperse its plants to avoid concentrated needs of resources such as water and energy. Nevertheless, such advanced foundries will undoubtedly require qualified professional staff in sufficient numbers which tends to favor locations close to metropolitan centers rather than in remote places. Last but not least, having this industry relocated in Europe will likely come together with strict environmental legislation and constraints, which strongly supports the need for an in-depth monitoring of the environmental impacts. Anticipation is therefore critical to have a better understanding of the environmental burden generated by semiconductor manufacturing. Finally, the growth rate of the semiconductor industry observed in Taiwan, in particular at TSMC, should also question the technological choice of CMOS technology nodes to be manufactured. More research has to be done on this matter. 

\subsection{Limitations of this Study}
\label{subsec:limits}

This study focuses mostly on companies responsible for the front-end IC processing of the chip production chain.
Further analyses should however also include remaining activities such as printed circuit board (PCB) manufacturing, the back-end processing and the manufacturing of other electronic components. The activities of the companies surveyed are sometimes unclear and it is possible that some companies fall outside the original scope but still remain within the Electronics Components Manufacturers sector and therefore do not affect the results or the purpose of this study. Further research should include more component types and a more extensive production chain, i.e., design, packaging, processing, testing.
For the same reasons, the extrapolation done in Section~\ref{subsec:comparison} assumes that the GHG emissions of all Taiwanese ECMs are due to the generation of electricity, but it is impossible to verify the soundness of this hypothesis without reviewing all companies of the sub-sector.

Moreover, only 16 companies out of the 60 surveyed issue CSR reports with useful data, while more than 100 companies could be reviewed if all production phases were included. Obtaining data from all semiconductor manufacturers present in Taiwan is still problematic for two reasons: either the environmental data is simply not reported, or the data provided is not location-based (this is the case for AU Optronics, Macronix, Global Foundries, Micron, Merck, Delta, Lite-On Semiconductor, Panjit or Foxsemicon). The disclosure of data specific to their Taiwanese operations would significantly increase the coverage of this study. We also observed that data present in older CSR reports are sometimes changed or updated in more recent editions. Such changes raise fundamental questions about the meaning or validity of the data contained in the CSR reports. In addition, some data among the ones recovered do not report values over a sufficient number of years or report values in a non standardized format. These limitations illustrate the important improvements in CSR reporting that could be achieved to allow a more effective processing and understanding of the environmental impacts of companies. Yet, in a field that suffers from a chronic lack of open and accessible data, it is important to note that this study is possible only thanks to the open data policies set up in Taiwan. Encouraging transparency and open-data policies could foster higher quality research and allow a more accurate identification of the environmental footprint of industrial activities.

Even though this study includes 16 companies of various sizes and covers half of the electricity consumption of the sub-sector, TSMC largely influences the results. On average, this company represents half of all studied environmental impacts. Further, most of the observed growth in the indicators originate from TSMC. Due to high-level reporting of the environmental data in the CSR reports of the company, it is however difficult to clearly identify whether these increases are due to yield changes, technological changes (e.g., the use of EUV for nodes below 10 nm) or the increase of production volumes. More granular data and interviews would be required to obtain a more precise picture. 
Nevertheless, TSMC is clearly the world leader in the manufacturing of sub-10nm nodes \cite{varas2021strengthening}, which totally justifies its inclusion in this study. When taking TSMC out of the analysis, we note that the rest of the companies show a slower growth rate for the three environmental indicators.

Regarding the GHG emissions, the reported CSR data are usually well detailed for the Scopes 1 and 2. Nevertheless, few companies report the use of self-produced renewable energy on their facilities. In the case of TSMC, this represents only 6.8\% of the company's electricity consumption in 2019\cite{TSMC-2020} but the company increased its renewable energy self-consumption by 818\% in 5 years, up to 918~GWh. Beyond Scopes 1 and 2, the general lack of data and standardized reporting regarding the Scope 3 hinder an in-depth understanding of the emissions throughout the entire value chain, although this could hide significant impacts. For instance, TSMC indicates that it has increased its Scope 3 emissions by 63\% between 2015 and 2020 \cite{TSMC-2020}.

\section{Conclusion} \label{sec:conclusion}

In this study, we show that the absolute environmental impacts of the electronic components manufacturing industry in Taiwan are on the rise over the period 2015-2020. An increasing trend is observed for each indicator, with a CAGR of 8.8\%, 8.9\%, 7.5\%, and 6.1\% for final energy consumption, electricity consumption, GHG emissions, and water consumption, respectively. It is clear that total absolute environmental impacts keep increasing on the island, despite the (relative) efficiency improvements claimed by the industry and the strong efforts to develop state-of-the-art manufacturing technologies.

Through a scenario analysis, we also highlight the conflicting trends between the ECMs' growing carbon footprint and the carbon neutrality roadmap outlined by the Taiwanese government up to 2050. As these companies are a keystone of the island's industrial, commercial and geopolitical policies, it raises a serious challenge for Taiwan to succeed in its ecological transition. To meet its targets, Taiwan could rely on a better efficiency from new production processes and a heavy decarbonation of its electricity supply. However, even TSMC's footprint indicates that the most advanced machinery might not be sufficient to curb the total GHG emissions in the coming years, especially if both the production volumes and the environmental impacts of advanced CMOS technology nodes continue to grow \cite{pirson2022essderc}. A quick switch to low-carbon energy sources still seems out of sight for an economy that relies mainly on imported carbon-intensive energy sources such as coal, crude oil and LNG. Other options could be to cap production volumes and number of plants, maintaining the most resource-efficient processes and rapidly stabilizing or reducing energy consumption in absolute values. Yet, this would have direct economic effects in conflict with the objective of the government who clearly wants to maintain and strengthen the economic outputs. Furthermore, the electronics industry is one of the major geopolitical assets of the island, referred as the \textit{silicon shield}, for its use as a deterrent against possible Chinese ambitions \cite{addison2001silicon}.

Overall, Taiwan's carbon neutrality roadmap will have to deal with an industry that is likely to increase its GHG emissions by 3.5 to 7.5\% per year. Facing such trends, it is unlikely than Taiwan will be able to both maintain its current industrial development and its carbon neutrality roadmap. As a consequence, Taiwanese semiconductors components will remain with a relatively high carbon footprint for a long time. In addition, the electronics industry itself is facing the increasing pressure of its environment. The recent drought caused by the absence of typhoons in 2020 and the water restrictions put in place in 2021 can be an early sign of the incoming risks for this industry.

Although this study already gathers enough data to cover a significant part of the ECM industry on the island, i.e., 56 to 70\%, further work is needed to extend the scope, improve the accuracy, and include more companies. Yet, this work can be seen as a first attempt to lay the foundations for an urgent work of monitoring. We also point out the necessity for ECMs' CSR reports to provide more details in the reporting of their environmental impacts. The open-data culture in Taiwan is a real opportunity to develop an in-depth understanding of ECMs' environmental footprint by answering these two challenges.

\section*{Acknowledgments}

This work was supported by the Fonds européen de développement régional (FEDER) and the Wallonia within the Wallonie-2020.EU program. The authors would like to thank the ECS group members at UCLouvain for their proofreading.
\section*{Supplementary material}

The supplementary material includes all the data and assumptions used in this study.

\bibliographystyle{ieeetr}

\bibliography{bibliography.bib}{}

\begin{thebibliography}{10}

\bibitem{itri2022}
{Yang, Ray (Consulting Director)}, ``{Overview of Taiwan Semiconductor Industry
  and National R\&D Initiatives - Presentation at the 22nd Taiwan-Belgium Joint
  Business Council Meeting, ITRI Industry - Science and Technology
  International Strategy Center},'' 2022.
\newblock (Webinar given on May, 5th 2022).

\bibitem{MOF-2021}
{Ministry of Finances}, ``{Annual External Trade Report in 2020},'' 2021.
\newblock
  \url{http://service.mof.gov.tw/public/Data/statistic/bulletin/109/2020.pdf}
  (accessed January 2022).

\bibitem{chou2019conundrums}
K.~Chou, D.~Walther, and H.~Liou, ``{The Conundrums of Sustainability: Carbon
  Emissions and Electricity Consumption in the Electronics and Petrochemical
  Industries in Taiwan},'' {\em Sustainability}, vol.~11, no.~20, p.~5664,
  2019.
\newblock \url{https://doi.org/10.3390/su11205664}.

\bibitem{boyd2012life}
S.~B. Boyd, {\em Life-cycle assessment of semiconductors}.
\newblock Springer Science \& Business Media, 2012.
\newblock \url{https://doi.org/10.1007/978-1-4419-9988-7}.

\bibitem{bardon2020dtco}
M.~G. Bardon, P.~Wuytens, L.-{\AA}. Ragnarsson, G.~Mirabelli, D.~Jang,
  G.~Willems, A.~Mallik, A.~Spessot, J.~Ryckaert, and B.~Parvais, ``{DTCO
  including sustainability: Power-performance-area-cost-environmental score
  (PPACE) analysis for logic technologies},'' in {\em 2020 IEEE International
  Electron Devices Meeting (IEDM)}, pp.~41--4, IEEE, 2020.
\newblock \url{https://doi.org/10.1109/IEDM13553.2020.9372004}.

\bibitem{moreau2021could}
N.~Moreau, T.~Pirson, G.~Le~Brun, T.~Delhaye, G.~Sandu, A.~Paris, D.~Bol, and
  J.-P. Raskin, ``{Could Unsustainable Electronics Support Sustainability?},''
  {\em Sustainability}, vol.~13, no.~12, p.~6541, 2021.
\newblock \url{https://doi.org/10.3390/su13126541}.

\bibitem{voas2021scarcity}
J.~Voas, N.~Kshetri, and J.~F. DeFranco, ``{Scarcity and Global Insecurity: The
  Semiconductor Shortage},'' {\em IT Professional}, vol.~23, no.~5, pp.~78--82,
  2021.
\newblock \url{https://doi.org/10.1109/MITP.2021.3105248}.

\bibitem{chips-act-2022}
{European Commission}, ``{A Chips Act For Europe: Communication From The
  Commission To The European Parliament, The Council, The European Economic And
  Social Committee And The Committee Of The Regions},'' 2022.
\newblock \url{https://ec.europa.eu/newsroom/dae/redirection/document/83086}
  (accessed May 2022).

\bibitem{acm2022}
{ACM Technology Policy Office}, ``{Comments Of The ACM Europe Technology Policy
  Committee on The Proposal For A Regulation Establishing A framework Of
  Measures For Strengthening europe's Semiconductor Ecosystem},'' 2022.
\newblock
  \url{https://europe.acm.org/binaries/content/assets/public-policy/europetpc-chips-act-comments-09may22.pdf}
  (accessed June 2022).

\bibitem{pirson2022essderc}
T.~Pirson, T.~Delhaye, A.~Pip, G.~Le~Brun, J.-P.~R. Raskin, and D.~Bol, ``{The
  Environmental Footprint of IC Production: Meta-Analysis and Historical
  Trends},'' {\em IEEE ESSDERC 2022}, 2022.

\bibitem{varas2021strengthening}
A.~Varas, R.~Varadarajan, J.~Goodrich, and F.~Yinug, ``{Strengthening The
  Global Semiconductor Supply Chain In An Uncertain Era},'' {\em Boston
  Consulting Group and Semiconductor Industry Association}, 2021.
\newblock
  \href{https://www.semiconductors.org/wp-content/uploads/2021/05/BCG-x-SIA-Strengthening-the-Global-Semiconductor-Value-Chain-April-2021_1.pdf}{https://www.semiconductors.org/wp-content/uploads/2021/05/BCG-x-SIA-Strengthening-the-Global-Semiconductor-Value-Chain-April-2021\_1.pdf}
  (accessed May 2022).

\bibitem{krishnan2008hybrid}
N.~Krishnan, S.~Boyd, A.~Somani, S.~Raoux, D.~Clark, and D.~Dornfeld, ``A
  hybrid life cycle inventory of nano-scale semiconductor manufacturing,'' {\em
  Environmental science \& technology}, vol.~42, no.~8, pp.~3069--3075, 2008.
\newblock \url{https://doi.org/10.1021/es071174k}.

\bibitem{hu2003power}
S.-C. Hu and Y.~Chuah, ``{Power consumption of semiconductor fabs in Taiwan},''
  {\em Energy}, vol.~28, no.~8, pp.~895--907, 2003.
\newblock \url{https://doi.org/10.1016/S0360-5442(03)00008-2}.

\bibitem{liu2010life}
C.~Liu, S.~J. Lin, and C.~Lewis, ``{Life cycle assessment of DRAM in Taiwan's
  semiconductor industry},'' {\em Journal of Cleaner Production}, vol.~18,
  no.~5, pp.~419--425, 2010.
\newblock \url{https://doi.org/10.1016/j.jclepro.2009.10.004}.

\bibitem{hu2016assessing}
A.~H. Hu, L.~H. Huang, and Y.-L. Chang, ``{Assessing corporate sustainability
  of the ICT sector in Taiwan on the basis of UN Sustainable Development
  Goals},'' in {\em 2016 Electronics goes green 2016+(EGG)}, pp.~1--6, IEEE,
  2016.
\newblock \url{https://doi.org/10.1109/EGG.2016.7829866}.

\bibitem{lin2018sustainability}
F.~Lin, S.-W. Lin, and W.-M. Lu, ``{Sustainability assessment of Taiwan’s
  semiconductor industry: A new hybrid model using combined analytic hierarchy
  process and two-stage additive network data envelopment analysis},'' {\em
  Sustainability}, vol.~10, no.~11, p.~4070, 2018.
\newblock \url{https://doi.org/10.3390/su10114070}.

\bibitem{hsu2017investigating}
C.-W. Hsu and D.-S. Chang, ``{Investigating critical organizational factors
  toward sustainability index: Insights from the Taiwanese electronics
  industry},'' {\em Business Ethics: A European Review}, vol.~26, no.~4,
  pp.~468--479, 2017.
\newblock \url{https://doi.org/10.1111/beer.12154}.

\bibitem{huang2020planetary}
L.~H. Huang, A.~H. Hu, and C.-H. Kuo, ``{Planetary boundary downscaling for
  absolute environmental sustainability assessment—Case study of Taiwan},''
  {\em Ecological Indicators}, vol.~114, p.~106339, 2020.
\newblock \url{https://doi.org/10.1016/j.ecolind.2020.106339}.

\bibitem{bjorn2016proposal}
A.~Bj{\o}rn, M.~Margni, P.-O. Roy, C.~Bulle, and M.~Z. Hauschild, ``{A proposal
  to measure absolute environmental sustainability in life cycle assessment},''
  {\em Ecological Indicators}, vol.~63, pp.~1--13, 2016.
\newblock \url{https://doi.org/10.1016/j.ecolind.2015.11.046}.

\bibitem{Chiu-2011}
C.~HM, ``{The Dark Side of Silicon Island: High-Tech Pollution and the
  Environmental Movement in Taiwan},'' 2011.
\newblock Capitalism Nature Socialism, 22(1): 40-57
  \url{https://doi.org/10.1080/10455752.2010.546647}.

\bibitem{prakash2013schaffung}
S.~Prakash {\em et~al.}, ``{Schaffung einer Datenbasis zur Ermittlung
  {\"o}kologischer Wirkungen der Produkte der Informations-und
  Kommunikationstechnik (IKT)},'' 2013.
\newblock
  \url{https://www.umweltbundesamt.de/sites/default/files/medien/378/publikationen/texte_82_2013_janssen_informationstechnik_teil_c.pdf}.

\bibitem{ranganathan2004greenhouse}
J.~Ranganathan, L.~Corbier, P.~Bhatia, S.~Schmitz, P.~Gage, and K.~Oren, ``{The
  greenhouse gas protocol: A corporate accounting and reporting standard
  (revised edition)},'' 2004.

\bibitem{TSIA-2021}
T.~S.~I. Association, ``Member directory,'' 2021.
\newblock \url{https://www.tsia.org.tw/EN/MemberList?nodeID=59} (accessed
  January 2022).

\bibitem{SEMI-2021}
SEMI, ``Member directory,'' 2021.
\newblock \url{https://www.semi.org/eu/resources/memberdirectory?
  search=&category\%5B586\%5D=586&count ry\%5B681\%5D=681&az} (accessed January
  2022).

\bibitem{gri-home}
{Global Reporting Index}, ``Setting the agenda for the future,'' 2022.
\newblock \url{https://www.globalreporting.org/} (accessed May 2022).

\bibitem{tsia-overview}
{TSIA}, ``{Overview on the Taiwanese Semiconductor Industry},'' 2021.
\newblock \url{https://www.tsia.org.tw/PublOverview?nodeID=30} (accessed May
  2022).

\bibitem{IEA-Taipei}
IEA, ``{Electricity generation by source, Chinese Taipei 1990-2020},'' 2022.
\newblock \url{https://www.iea.org/countries/chinese-taipei} (accessed May
  2022).

\bibitem{BoE-2021}
{Bureau of Energy (Taipei)}, ``{Energy Statistics Handbook 2020},'' 2021.

\bibitem{NDC-2021}
N.~D. Council, ``{Taiwan Power Company purchased electricity (energy type) and
  sold electricity (use type) over the years},'' 2020.
\newblock \url{https://data.gov.tw/dataset/35392} (accessed January 2022).

\bibitem{TSMC-2020}
TSMC, ``{Corporate Social Responsibility Report (June, 2019)},'' 2020.

\bibitem{UMC-2020}
UMC, ``{Corporate Social Responsibility Report (July 2019)},'' 2020.

\bibitem{Innolux-2020}
Innolux, ``{Corporate Social Responsibility Report 2019},'' 2020.

\bibitem{MOEA-2021}
{Ministry of Economic Affairs}, ``{Industrial Production, Shipment \& Inventory
  Index Statistics Surveystics, Indexes of Industrial Production Manufacture of
  Electronic Parts and Components},'' 2021.
\newblock
  \url{https://dmz26.moea.gov.tw/GMWeb/investigate/InvestigateDA.aspx?lang=E}
  (accessed December 2021).

\bibitem{bol2021moore}
D.~Bol, T.~Pirson, and R.~Dekimpe, ``{Moore's Law and ICT Innovation in the
  Anthropocene},'' in {\em 2021 Design, Automation \& Test in Europe Conference
  \& Exhibition (DATE)}, pp.~19--24, IEEE, 2021.
\newblock \url{https://doi.org/10.23919/DATE51398.2021.9474110}.

\bibitem{pop-density-taiwan}
{Statistical Bureau}, ``{Population density (National Statistics, Republic of
  China)},'' 2022.
\newblock \url{https://eng.stat.gov.tw/point.asp?index=9} (accessed May 2022).

\bibitem{gao2021review}
A.~M.-Z. Gao, C.-H. Huang, J.-C. Lin, and W.-N. Su, ``{Review of recent
  offshore wind power strategy in Taiwan: Onshore wind power comparison},''
  {\em Energy Strategy Reviews}, vol.~38, p.~100747, 2021.
\newblock \url{https://doi.org/10.1016/j.esr.2021.100747}.

\bibitem{tsai2021trend}
W.-T. Tsai, ``{Trend analysis of Taiwan’s greenhouse gas emissions from the
  energy sector and its mitigation strategies and promotion actions},'' {\em
  Atmosphere}, vol.~12, no.~7, p.~859, 2021.
\newblock \url{https://doi.org/10.3390/atmos12070859}.

\bibitem{reuters}
{Reuters}, ``{Chipmakers in drought-hit Taiwan order water trucks to prepare
  for 'the worst'}.''
\newblock
  \url{https://www.reuters.com/article/us-taiwan-drought-semiconductors-idUSKBN2AO0G3},
  accessed May 2022.

\bibitem{unepGAP-2019}
UNEP, ``{Emissions Gap Report 2019. Executive summary. United Nations
  Environment Programme, Nairobi.},'' 2019.
\newblock
  \url{https://wedocs.unep.org/bitstream/handle/20.500.11822/30798/EGR19ESEN.pdf?sequence=13}
  (accessed May 2022).

\bibitem{freitag2021real}
C.~Freitag, M.~Berners-Lee, K.~Widdicks, B.~Knowles, G.~S. Blair, and
  A.~Friday, ``{The real climate and transformative impact of ICT: A critique
  of estimates, trends, and regulations},'' {\em Patterns}, vol.~2, no.~9,
  p.~100340, 2021.
\newblock \url{https://doi.org/10.1016/j.patter.2021.100340}.

\bibitem{Pirson2021}
T.~Pirson and D.~Bol, ``{Assessing the embodied carbon footprint of IoT edge
  devices with a bottom-up life-cycle approach},'' {\em Journal of Cleaner
  Production}, vol.~322, p.~128966, 2021.
\newblock \url{https://doi.org/10.1016/j.jclepro.2021.128966}.

\bibitem{unruh2000understanding}
G.~C. Unruh, ``Understanding carbon lock-in,'' {\em Energy policy}, vol.~28,
  no.~12, pp.~817--830, 2000.
\newblock \url{https://doi.org/10.1016/S0301-4215(00)00070-7}.

\bibitem{addison2001silicon}
C.~Addison, {\em {Silicon Shield: Taiwan's Protection Against Chinese Attack}}.
\newblock Fusion Press, 2001.

\bibitem{MOEA-incentive}
{MOEA}, ``{Taiwan Key Innovative Industry - Semiconductors}.''
\newblock
  \url{https://www.roc-taiwan.org/uploads/sites/30/2018/03/Semiconductors.pdf}
  (accessed June 2022).

\bibitem{WSJ}
{Yang Jie}, ``{TSMC to Invest Up to 44 US Billion in 2022 to Beef Up Chip
  Production}.''
\newblock
  \url{https://www.roc-taiwan.org/uploads/sites/30/2018/03/Semiconductors.pdf}
  (accessed June 2022).

\end{thebibliography}

\clearpage
\onecolumn




\section*{Supplementary material}

# VII. Supplementary material

## Listing of electronic components manufacturers (ECMs) in Taiwan

| Company name | Fab location in Taiwan | CSR Data available | Taiwan-related data available | Within scope | Included in study | Specialization |
|---|---|---|---|---|---|---|
| **Included** | | | | | | |
| TSMC | Fab2, Fab3, Fab5, Fab12, Fab Yes | Yes | Yes | Yes | Yes | Logic IC, 5nm to 450nm; 6-inch to 12-inch |
| UMC | FabBA, Fab8AB-N (Hsinchu); Yes | Yes | Yes | Yes | Yes | 200nm IC / 300nm IC; 8-inch / 12-inch |
| Powerchip | FabP1, FabP2, FabP3, FabBA Yes | Yes | Yes | Yes | Yes | Power management IC / CMOS Image Sensor / Integrated Memory Chip / Discrete Devices for IGBT and Power MOS / LCD Driver IC (Opto) / SLC-NAND Flash / Low Power DRAM; 12-inch |
| Vanguard | Fab1, Fab2 (Hsinchu); Fab3 (Yes | Yes | Yes | Yes | Yes | Display drivers IC / Power management IC / Discrete components; 8-inch |
| Global Wafers | Hsinchu | Yes | Yes | Yes | Yes | 3-inch to 12-inch silicon wafers |
| Nanya | Fab, Fab2 (Linkou) ; Fab3 (New Yes | Yes | Yes | Yes | Yes | DRAM |
| Winbond | Memory Product Foundry (Taic Yes | Yes | Yes | Yes | Yes | Flash Memory / Speciality DRAM / Mobile DRAM; 12-inch |
| Win Semiconductors | Fab A, Fab B, Fab C (Taoyuan) Yes | Yes | Yes | Yes | Yes | GaAs devices |
| Epistar | FabF1 (Longtan) ; FabA1, N2, Yes | Yes | Yes | Yes | Yes | LED (Opto) |
| Unimicron | Shanying, Luzhu, Hejiang, Zho Yes | Yes | Yes | Yes | Yes | PCB / HDI / Carrier board / RF / IC Testing |
| Innolux | Zhunan, Tainan | Yes | Yes | Yes | Yes | TFT-LCD |
| Nuvoton | Hsinchu plant | Yes | Yes | Yes (partially) | Yes | General Purpose IC / Microcontrollers; 6-inch (Wafer mask manufacturing / testing / packaging) |
| Lextar | T01 (Hsinchu) | Yes | Yes | Yes | Yes | LED (Opto) |
| Everlight | Yuanli Plant, Shulin Plant, Tong Yes | Yes | Yes | Yes | Yes | LED (Opto) |
| Opbtech | Hsinchu | Yes | Yes | Yes | Yes | LED (Opto) |
| Wafer Works | Yangmei, Longtan | Yes | Yes | Yes | Yes | Wafer manufacturing |
| **Not included** | | | | | | |
| Siliconware Precision Industries | Dafong, Chungshan, Zhongke | Yes | Yes | No | No | IC packaging, processing, testing |
| Foxsemicon Integrated Technology | Hsinchu | Yes | No (not enough data) | | No | |
| Orient Semiconductor Electronics | Kaohsiung | Yes | No (not enough data) | | No | |
| Macronix | Fab2, Fab5, Fab1 (Hsinchu) | Yes | No (not enough data) | | No | |
| AUO | Fab L3D, L5D (Gueishan) ; Fab Yes | Yes | No | | No | |
| Global Foundries | USA, Germany, Singapore | Yes | No | | No | |
| Micron Memory Taiwan | Micron Memory (Taichung) / Fa Yes | Yes | No | | No | |
| Merck Performance Materials | Kaohsiung | Yes | No | | No | |
| Panjit | Kaohsiung | Yes | No | | No | |
| Lite-On Semiconductor | Plant 1 (Keelung) / Plant 2 (Hsi Yes | Yes | No | | No | |
| Delta Electronics | | Yes | No | | No | |
| Advanced Semiconductor Engineering Inc. (ASE Group +SPIL) | | Yes | No | | No | |
| Tatung Corp | | Yes | No | | No | |
| Acer | | Yes | No | | No | |
| Elan Microelectronics | Hsinchu, Zhonghe, Tainan, Kac Yes | Yes | No | | No | |
| Formosa Sumco Technology | Kuang-Fu, Li-Hsin Prosperity, F Yes | Yes | No | | No | |
| Chipbond | Tainan | No | | | No | |
| Gudeng Precision | Device Foundry, 6A, 6B (Hsinci No | No | | | No | |
| Epistil | Fab4 (Taichung) | No | | | No | |
| ProMOS | Lije, I-Lan | No | | | No | |
| Taiwan Semi | Hsinchu | No | | | No | |
| Arima Optoelectronics | Tainan | No | | | No | |
| AWSC | Taoyuan | No | | | No | |
| SAN CHIH Semiconductor | Wuxi Creative Sensor (Taipei) | No | | | No | |
| Creative Sensor Inc. | Headquarters Phase 1 (Hsinch No | No | | | No | |
| Visera Technologies (Part of TSMC) | | No | | | No | |
| Epistl | | No | | | No | |
| Creating Nano | | No | | | No | |
| Epistl-Precision | | No | | | No | |
| Sun Yuan Technology | | No | | | No | |
| U-Can Dynatex | | No | | | No | |
| Ofuna Technology | | No | | | No | |
| Cin Phown Technology | | No | | | No | |
| Davicom Semiconductor | | No | | | No | |
| Holtek Semiconductor | | No | | | No | |
| Realtek Semiconductor | | No | | | No | |
| WiseChip Semiconductor | | No | | | No | |
| Unikorn Semi | Hsinchu | No | | | No | |
| Sigurd Microelectronics | Peishing, Chungsheng, Hukou (No | No | | | No | |
| Mosel Vitelic | | No | | | No | |
| Kneron | | Not relevant | | | No | |
| Green Energy Technology | | No | | | No | Solar panels |
| Sino-American Silicon Products | | No | | | No | Crystalline Wafer, Solar cell, Solar module |
| Hexawave | | No | | | No | |
| COUNT | 60 | COUNT 33 | COUNT 17 | COUNT 16 | COUNT 16 | |

*Notes*

* The term "Electronic Components manufacturers" (ECM) can include: semiconductor manufacturing, passive electronic component manufacturing, optoelectronic materials and components, other electronic component manufacturing.

*Sources*

https://www.tsia.org.tw/EN/MemberList?nodeID=59 (Criterion : Category « Manufacturing »)
https://www.semi.org/eu/resources/member-directory?search=&category=%5B858%5D=5868&country%5B881%5D=881&az (Criterion : Category « Semiconductor », Semiconductor, Photonics » + Category « Device manufacturing » + Country «Taiwan »)
https://database.globalreporting.org/search/ (Criterion: Keyword « Semi » + « Country «Taiwan»)
https://en.wikipedia.org/wiki/List_of_semiconductor_fabrication_plants (Criterion : Plant location «Taiwan»)

## Summary of results

| Year | 2015 | 2016 | 2017 | 2018 | 2019 | 2020 | CAGR |
|---|---|---|---|---|---|---|---|
| **GHG emissions (Unit : metric tons CO2e)** | | | | | | | |
| **THIS STUDY – GHG Emissions (Sc1+2)** | 13 085 218 | 14 158 264 | 15 712 523 | 16 969 309 | 17 200 259 | 18 770 908 | 7,5% |
| Industry GHG emissions | 126 330 000 | 127 590 000 | 131 210 000 | 132 950 000 | 127 070 000 | ND | |
| THIS STUDY – share of Industry GHG emissions (%) | 10,4 | 11,1 | 12,0 | 12,8 | 13,5 | ND | |
| National GHG Emissions | 260 200 000 | 264 740 000 | 271 220 000 | 268 850 000 | 260 400 000 | ND | 0,0% |
| THIS STUDY – share of National GHG emissions (%) | 5,0 | 5,3 | 5,8 | 6,3 | 6,6 | ND | |
| THIS STUDY (SECTOR) - share of National GHG emissions (%) | 8,1 | 8,6 | 8,8 | 9,0 | 11,7 | ND | |
| **Final energy consumption (Unit : GWh)** | | | | | | | |
| **THIS STUDY – final energy consumption** | 18 947 | 20 495 | 25 131 | 26 515 | 27 904 | 31 403 | 8,8% |
| Industry final energy consumption | 294 101 | 297 016 | 297 418 | 302 286 | 299 450 | 312 579 | 1,2% |
| THIS STUDY – share of Industry final energy consumption (%) | 6,4 | 6,9 | 8,4 | 8,8 | 9,3 | 10,0 | |
| National final energy consumption | 774 059 | 780 075 | 775 570 | 788 470 | 761 440 | 769 250 | -0,1% |
| THIS STUDY – share of national final energy consumption (%) | 2,4 | 2,6 | 3,2 | 3,4 | 3,7 | 4,1 | |
| THIS STUDY (SECTOR) - share of National final energy consumption (%) | 3,9 | 4,2 | 4,9 | 4,8 | 6,5 | 7,0 | |
| **Electricity consumption (Unit : GWh)** | | | | | | | |
| **THIS STUDY – electricity consumption** | 19 217 | 20 309 | 22 807 | 25 131 | 26 505 | 29 421 | 8,9% |
| Industry electricity consumption | 134 700 | 136 890 | 141 111 | 148 961 | 147 675 | 150 742 | 2,3% |
| THIS STUDY – share of Industry electricity consumption (%) | 14,3 | 14,8 | 16,2 | 16,9 | 17,9 | 19,5 | |
| National electricity consumption | 250 019 | 255 420 | 261 394 | 266 568 | 265 720 | 271 247 | 1,6% |
| THIS STUDY – share of national electricity consumption (%) | 7,7 | 8,0 | 8,7 | 9,4 | 10,0 | 10,8 | |
| THIS STUDY (SECTOR) - share of National electricity consumption (%) | 12,3 | 12,8 | 13,3 | 13,4 | 17,7 | 18,7 | |
| **Water consumption (Unit : m3)** | | | | | | | |
| **THIS STUDY – water consumption** | 107 207 300 | 105 927 409 | 117 015 590 | 123 979 533 | 129 511 532 | 144 040 335 | 6,1% |
| Industry water consumption | 1 601 000 000 | 1 629 000 000 | 1 654 010 000 | 1 668 000 000 | 1 671 350 000 | 1 803 190 000 | 2,4% |
| THIS STUDY – share of Industry water consumption (%) | 6,7 | 6,5 | 7,1 | 7,4 | 7,7 | 8,0 | |
| National water consumption | 16 025 000 000 | 16 546 000 000 | 16 645 000 000 | 16 713 000 000 | 16 739 280 000 | 16 675 930 000 | 0,8% |
| THIS STUDY – share of National water consumption (%) | 0,7 | 0,6 | 0,7 | 0,7 | 0,8 | 0,9 | |

## Indexes of environmental impacts and manufacture of electronic parts and components

| 2016=100 | 2015 | 2016 | 2017 | 2018 | 2019 | 2020 | CAGR |
|---|---|---|---|---|---|---|---|
| THIS STUDY – GHG Emissions (Sc1+2) | 92,4 | 100,0 | 111,0 | 119,9 | 121,5 | 132,6 | 7,48% |
| THIS STUDY – Electricity consumption | 94,6 | 100,0 | 112,3 | 123,7 | 130,5 | 144,9 | 8,89% |
| THIS STUDY – Water consumption | 101,2 | 100,0 | 110,5 | 117,0 | 122,3 | 136,0 | 6,08% |
| Manufacture of Integrated Circuits | 88,1 | 100,0 | 109,4 | 119,0 | 120,7 | 153,9 | 11,80% |

## Study representativeness of Taiwanase ECM industry

| Year | 2015 | 2016 | 2017 | 2018 | 2019 | 2020 | (min) | (max) | (average) |
|---|---|---|---|---|---|---|---|---|---|
| Autoproducers share of national electricity generation | 16,40 | 16,44 | 16,17 | 15,22 | 15,53 | 16,00 | | | |
| Independent power producers share of national electricity generation | 15,55 | 14,97 | 14,37 | 15,81 | 15,93 | 15,78 | | | |
| Taipower share of national electricity generation | 68,05 | 68,59 | 69,46 | 68,97 | 68,54 | 68,22 | | | |
| Share of Taipower ECM electricity consumption | ND | 62,34 | 65,82 | 70,23 | 56,31 | 57,94 | | | |
| **Representativness of study based on Taipower data** | ND | 62,3 | 65,8 | 70,2 | 56,3 | 57,9 | 56 | 70 | 62,5 |

*Sources*
MOEA, Energy Statistics Handbook 2020 p.83-84 ; Taipower Open Data, 台灣電力公司_
歷年售電量(用途別)) (https://data.gov.tw/dataset/35392)

https://www.moeaboe.gov.tw/ECW/main/content/wHandMenuFile.ashx?file_id=1411
https://www.moeaboe.gov.tw/ECW/english/content/ContentLink.aspx?menu_id=1540

Taipower produces about 70% of the electricity on the island but it is the main electricity vendor for ECMs. We therefore consider Taipower's electricity to represent 100% of the sub-sector.



# Greenhouse Gases Emissions of selected electronic components manufacturers

| Unit: metric tons CO2e | Greenhouse Gases Emissions (Scope 1) | | | | | | Greenhouse Gases Emissions (Scope 2) | | | | | | Total Greenhouse Gases Emissions (Scope 1+2) | | | | | | Notes | Source |
|---|---|---|---|---|---|---|---|---|---|---|---|---|---|---|---|---|---|---|---|---|
| Year | 2015 | 2016 | 2017 | 2018 | 2019 | 2020 | Diff 2015/2020 (%) | 2015 | 2016 | 2017 | 2018 | 2019 | 2020 | Diff 2015/2020 (%) | 2015 | 2016 | 2017 | 2018 | 2019 | 2020 | Diff 2015/2020 (%) | | |
| TSMC | 1,561,662 | 1,643,368 | 1,631,051 | 1,709,746 | 1,679,753 | 2,150,339 | 37 | 4,315,766 | 5,030,647 | 5,702,511 | 6,320,931 | 6,972,235 | 7,420,951 | 72 | 5,882,428 | 6,673,815 | 7,343,562 | 8,031,677 | 8,351,988 | 9,581,290 | 63 | 0.534 kg of CO2e / kWh | TSMC CSR Report 2019-2020 |
| Innolux | 428,000 | 440,000 | 425,000 | 348,000 | 305,000 | 320,000 | -25 | 2,398,000 | 2,277,000 | 2,040,000 | 2,078,000 | 2,473,000 | 3,089,000 | 29 | 2,816,000 | 2,711,000 | 2,465,000 | 2,347,000 | 3,104,000 | 3,392,000 | 20 | | Innolux CSR Report 2019-2020 |
| Unimicron | 15,542 | 16,850 | 16,500 | 18,540 | 17,635 | 20,847 | 31 | 436,653 | 415,152 | 424,828 | 520,861 | 519,638 | 509,522 | 28 | 452,195 | 431,992 | 440,984 | 534,401 | 537,003 | 540,969 | 20 | | Unimicron CSR Report 2019-2020 |
| UMC | 621,450 | 610,000 | 610,000 | 608,000 | 588,976 | 533,321 | -14 | 1,187,200 | 1,186,000 | 1,300,000 | 1,350,000 | 1,318,000 | 1,267,249 | 7 | 1,808,650 | 1,802,000 | 1,865,954 | 1,961,956 | 1,831,741 | 1,923,899 | 2 | 0.533 kg of CO2e / kWh | UMC CSR Report 2019-2020 |
| Powerchip | 45,460 | 59,684 | 89,978 | 107,183 | 249,134 | 107,951 | 132 | 426,066 | 430,177 | 565,158 | 502,656 | 540,034 | 665,138 | 43 | 471,354 | 495,681 | 655,670 | 659,749 | 643,815 | 673,030 | 43 | New: Fab12 in Singapore is not counted | Powerchip CSR Report 2019-2020 |
| Vanguard | 19,200 | 27,300 | 27,910 | 33,100 | 296,500 | 236,600 | 1,132 | 406,100 | 434,000 | 430,000 | 449,100 | 460,700 | 434,900 | 7 | 627,900 | 726,000 | 711,100 | 728,100 | 729,900 | 671,500 | 7 | | Vanguard CSR Report 2019-2020 |
| GlobalWafers | 225 | 1,160 | 1,737 | 1,634 | 1,863 | 1,970 | 1,475 | 10,695 | 191,205 | 156,865 | 202,115 | 174,994 | 169,827 | 1,472 | 10,950 | 192,389 | 156,885 | 202,115 | 174,994 | 169,827 | | | GlobalWafers CSR Report 2019-2020 |
| Nanya | 51,930 | 51,394 | 59,751 | 78,311 | 88,700 | 90,327 | 74 | 218,072 | 239,634 | 266,918 | 290,834 | 389,924 | 378,417 | 74 | 270,233 | 290,033 | 326,689 | 368,215 | 478,721 | 467,744 | 74 | | Nanya CSR Report 2015-2016-2019-2020 |
| Winbond | 51,725 | 51,673 | 62,318 | 64,510 | 65,749 | 52,271 | 1 | 208,181 | 209,959 | 251,458 | 285,898 | 290,598 | 283,105 | 31 | 259,906 | 261,632 | 313,771 | 331,408 | 331,317 | 331,317 | 31 | | Winbond CSR Report 2017-2018-2019-2020 |
| Quaker | 21,227 | 21,959 | 31,648 | 23,510 | 16,519 | 38,444 | 81 | 138,577 | 165,386 | 168,322 | 198,659 | 152,830 | 192,017 | 39 | 159,814 | 187,891 | 199,928 | 183,929 | 251,451 | 230,451 | 44 | Might include 1 small Chinese facility | Quaker CSR Report 2019-2020 |
| Vis Semiconductor | 41,627 | 48,743 | 54,616 | 64,447 | 69,342 | 70,860 | 70 | 70,281 | 82,874 | 86,662 | 101,283 | 104,515 | 114,853 | 56 | 112,918 | 131,617 | 151,078 | 165,730 | 170,463 | 191,843 | 70 | | Vis CSR Report 2019-2020 |
| Nuvoton | 32,990 | 34,486 | 39,910 | 36,951 | 41,900 | 36,108 | -10 | 41,100 | 40,196 | 40,100 | 41,600 | 38,055 | 37,725 | -8 | 73,090 | 74,762 | 80,010 | 79,200 | 80,700 | 73,833 | 4 | | Nuvoton CSR Report 2019-2020 |
| Lextar | 44,894 | 44,354 | 40,863 | 51,815 | 41,905 | 28,323 | -37 | 52,808 | 46,418 | 48,051 | 59,249 | 59,244 | 74,697 | 41 | 97,702 | 90,772 | 89,081 | 103,046 | 81,153 | 68,920 | -6 | | Lextar CSR Report 2015-2016-2017-2018-2019-2020 |
| Everlight | ND | ND | ND | ND | ND | 812 | | ND | ND | ND | ND | ND | 10,320 | | ND* | ND* | ND* | ND* | ND* | 10,878 | | Scopes not detailed | Everlight CSR Report 2019 |
| Optotech | ND* | ND* | 9,987 | ND* | ND* | ND* | | ND* | ND* | 20,782 | 39,109 | 47,125 | 41,946 | | 37,127 | 39,782 | 39,769 | 47,125 | 41,946 | 41,946 | 12 | | Optotech CSR Report 2019 |
| Walsin Works | 5,433 | 7,729 | 9,867 | 10,778 | 11,455 | 6,900 | 12 | 55,593 | 64,498 | 67,405 | 64,409 | 62,135 | 62,135 | 12 | 61,026 | 69,998 | 77,272 | 75,514 | 76,514 | 69,641 | 13 | Location-based data; might include 1 small Walsin Works CSR Report 2015-2018-2019-2020 |
| Subtotal | 3,120,897 | 3,227,469 | 3,547,018 | 3,546,065 | 3,848,065 | 3,846,176 | 23 | 9,932,791 | 10,763,050 | 12,328,011 | 13,472,036 | 13,942,642 | 14,988,817 | 51 | 13,092,216 | 14,190,254 | 15,712,923 | 17,017,561 | 17,262,208 | 19,770,909 | 43 | | |

(*) CSR data not contacted but didn't answer

## Greenhouse Gases emissions (GHG) in Taiwan

| Unit: metric tons CO2e | | | | | | | |
|---|---|---|---|---|---|---|---|
| Year | 2015 | 2016 | 2017 | 2018 | 2019 | 2020 | Diff 2015/2020 (%) | Source |
| National GHG Emissions | 280,030,000 | 284,740,000 | 271,720,000 | 288,690,000 | 285,410,000 | - | | NGER, Energy Statistics Handbook 2020 (p. 138); here, also here |
| Industry GHG emissions | 126,330,000 | 127,590,000 | 131,210,000 | 132,950,000 | 127,070,000 | - | | |
| ICT and semiconductor industry GHG emissions | 31,720,000 | 30,722,000 | 30,220,000 | 33,370,000 | 32,670,000 | - | | |
| Industry share of national GHG emissions | 45 | 45.2 | 48.4 | 46 | 44.5 | | | |
| THIS STUDY – Scope 1+2 Emissions of selected ECM | 13,092,216 | 14,190,254 | 15,712,923 | 16,969,306 | 17,240,276 | 17,200,209 | | |
| Share of industry GHG emissions | 10.4 | 11.1 | 12.0 | 12.8 | 13.6 | | | |
| Share of National GHG emissions | 5.0 | 5.3 | 5.8 | 6.3 | 6.0 | | | |
| **Other sectors** | | | | | | | | |
| Agriculture GHG emissions | 2,630,000 | 2,850,000 | 2,980,000 | 3,100,000 | 3,070,000 | | | |
| Transport GHG emissions | 37,040,000 | 38,160,000 | 37,620,000 | 36,790,000 | 37,000,000 | | | |
| Services GHG emissions | 29,770,000 | 29,160,000 | 30,950,000 | 28,440,000 | 27,740,000 | | | |
| Residential GHG emissions | 28,050,000 | 28,460,000 | 28,700,000 | 29,400,000 | 28,550,000 | | | |
| Industry GHG emissions | 126,330,000 | 127,590,000 | 131,210,000 | 132,950,000 | 127,070,000 | | | |
| Energy GHG emissions | 37,240,000 | 37,880,000 | 37,850,000 | 37,160,000 | 37,130,000 | | | |

| Greenhouse Gases emissions (GHG) in Taiwan | | | | | | |
|---|---|---|---|---|---|---|
| Unit: % | 2015 | 2016 | 2017 | 2018 | 2019 | 2020 |
| Share of scope 2 with respect to total (scope 1+2) | 75.9 | 75.2 | 76.5 | 76.4 | 76.2 | 76.1 |

## Water consumption of selected electronic components manufacturers

Unit : m3

| Year | 2015 | 2016 | 2017 | 2018 | 2019 | 2020 | Diff 2015/2020 (%) | Source |
|---|---|---|---|---|---|---|---|---|
| TSMC | 34 000 000 | 38 600 000 | 45 200 000 | 51 000 000 | 58 000 000 | 70 600 000 | 108 | TSMC CSR Report 2019-2021 |
| Innolux | 24 190 000 | 22 950 000 | 25 040 000 | 21 830 000 | 20 210 000 | 20 415 000 | -16 | Innolux CSR Report 2019-2020 |
| Unimicron | 18 752 000 | 13 061 106 | 13 140 492 | 14 614 868 | 15 306 000 | 15 007 898 | -20 | Unimicron CSR Report 2019-2020 |
| UMC | 13 830 000 | 14 456 000 | 14 900 000 | 14 910 000 | 14 810 000 | 15 500 000 | 12 | UMC CSR Report 2019-2020 |
| Powerchip | 3 678 669 | ND* | ND* | 1 451 979 | 1 497 624 | 1 560 296 | | Powerchip CSR Report 2019-2020 |
| Vanguard | 4 540 000 | 4 780 000 | 4 860 000 | 4 990 000 | 4 980 000 | 6 560 000 | 44 | Vanguard CSR Report 2019-2020 |
| Global Wafers | 545 097 | 3 639 324 | 3 664 318 | 5 469 000 | 4 960 000 | 4 324 000 | 693 | GlobalWafers CSR Report 2019-2020 |
| Nanya | 2 198 960 | 2 244 759 | 3 092 814 | 3 022 362 | 3 258 386 | 3 368 954 | 53 | Nanya CSR Report 2015-2016-2019-2020 |
| Winbond | 2 320 000 | 2 340 000 | 2 830 000 | 945 000 | 979 000 | 1 216 000 | -48 | Winbond CSR Report 2017-2018-2020 |
| Epistar | ND* | 450 000 | 420 000 | 410 000 | 380 000 | 370 000 | | Epistar CSR Report 2018-2020 |
| Win Semiconductors | ND* | ND* | 253 101 | 214 044 | 208 239 | 255 063 | | Win CSR Report 2018-2019-2020 |
| Nuvoton | 387 000 | 420 000 | 442 000 | 440 000 | 395 000 | 401 000 | 4 | Nuvoton CSR Report 2019-2020 |
| Lextar | ND* | ND* | ND* | 98 470 | 83 750 | 76 410 | | Lextar CSR Report 2016-2017-2018-2020 |
| Everlight | ND* | ND* | ND* | 994 490 | 874 770 | 849 600 | | Everlight CSR Report 2019-2020 |
| Optotech | 1 194 967 | 1 272 536 | 1 336 185 | 1 422 818 | 1 402 261 | 1 402 261 | 17 | Optotech CSR Report 2017-2019 |
| Wafer Works | 1 570 607 | 1 713 684 | 1 836 680 | 2 166 502 | 2 166 502 | 2 133 853 | 36 | Wafer Works CSR Report 2016-2018-2020 |
| **Subtotal** | **107 207 300** | **105 927 409** | **117 015 590** | **123 979 533** | **129 511 532** | **144 040 335** | | |
| All ECM except TSMC | 73 207 300 | 67 327 409 | 71 815 590 | 72 979 533 | 71 511 532 | 73 440 335 | 34 | |

ND = No Data

(*) : CSR department contacted but didn't answer

CSR reports are sometimes unclear about their reporting, it is difficult to understand if they're reporting water intake, water consumption including recycled water or their net water consumption. The numbers presented here should rather viewed as water intake or water footprint.

## Domestic water consumption in Taiwan

Unit : m3

| Year | 2015 | 2016 | 2017 | 2018 | 2019 | 2020 | Diff 2015/2020 (%) | Source |
|---|---|---|---|---|---|---|---|---|
| National water consumption | 16 025 000 000 | 16 546 000 000 | 16 645 000 000 | 16 713 000 000 | 16 739 280 000 | 16 675 930 000 | 4 | Utilization of Water Resources (2016-2019) https://www.wra.gov.tw/News.aspx?n=2953&sms=9084&_CSN=0 ; https://www.wra.gov.tw/News_Content.aspx?n=2945&s=7394 |
| Industry water consumption | 1 601 000 000 | 1 629 000 000 | 1 654 000 000 | 1 668 000 000 | 1 671 350 000 | 1 803 190 000 | 13 | |
| Industry share of national consumption | 10,0 | 9,8 | 9,9 | 10,0 | 10,0 | 10,8 | | |
| | | | Study perimeter | | | | | |
| THIS STUDY – Water consumption of s | 107 207 300 | 105 927 409 | 117 015 590 | 123 979 533 | 129 511 532 | 144 040 335 | 34,4 | |
| Share of Industry water consumption | 6,7 | 6,5 | 7,1 | 7,4 | 7,7 | 8,0 | | |
| Share of National water consumption | 0,7 | 0,6 | 0,7 | 0,7 | 0,8 | 0,9 | | |
| | | | Other sectors | | | | | |
| Agriculture water consumption | 10 497 740 000 | 11 086 030 000 | 11 200 240 000 | 11 890 000 000 | 11 363 500 000 | 11 592 470 000 | 10 | |
| Industry water consumption | 1 601 000 000 | 1 629 000 000 | 1 654 000 000 | 1 668 000 000 | 1 671 350 000 | 1 803 190 000 | 13 | |
| Domestic water consumption (household) | 3 142 230 000 | 3 183 410 000 | 3 147 140 000 | 3 155 810 000 | 3 185 510 000 | 3 280 270 000 | 4 | |

## Indexes of Industrial Production in Taiwan

*2016=100*

| Year | 2015 | 2016 | 2017 | 2018 | 2019 | 2020 | Notes | Sources |
|---|---|---|---|---|---|---|---|---|
| **Manufacture of Integrated Circuits** | 88,1 | 100,0 | 109,4 | 119,0 | 120,7 | 153,9 | *SCOPE* | MOEA |
| Manufacture of Discrete Devices | 96,0 | 100,0 | 109,3 | 128,3 | 124,3 | 130,0 | *SCOPE* | |
| Packaging and Testing of Semi-conductors | 99,8 | 100,0 | 103,8 | 108,3 | 116,6 | 123,8 | *OUT OF SCOPE* | |
| Manufacture of Electronic Passive Devices | 94,9 | 100,0 | 109,4 | 151,0 | 104,9 | 120,7 | *SCOPE* | |
| Manufacture of Bare Printed Circuit Boards | 99,6 | 100,0 | 108,7 | 109,7 | 107,4 | 120,1 | *OUT OF SCOPE* | |
| Manufacture of Liquid Crystal Panel and Components | 110,7 | 100,0 | 117,7 | 113,5 | 102,5 | 108,4 | *OUT OF SCOPE* | |
| Manufacture of Light Emitting Diodes (LED) | 118,6 | 100,0 | 91,2 | 84,0 | 69,0 | 64,5 | *SCOPE* | |
| Manufacture of Solar Cells | 110,1 | 100,0 | 83,4 | 74,0 | 46,2 | 44,5 | *OUT OF SCOPE* | |
| Manufacture of Other Optoelectronic Materials and Components | 105,8 | 100,0 | 88,8 | 84,5 | 83,3 | 82,5 | *SCOPE* | |
| Manufacture of Printed Circuit Assembly | 95,2 | 100,0 | 101,2 | 102,9 | 138,0 | 193,9 | *OUT OF SCOPE* | |
| Manufacture of Other Electronic Parts and Components (Not Elsewhere Classified) | 99,2 | 100,0 | 89,9 | 90,1 | 122,6 | 153,7 | *SCOPE* | |
| **Manufacture of Electronic Parts and Components (overall)** | **95,6** | **100,0** | **108,2** | **114,0** | **114,1** | **136,3** | | |

## Taiwan emissions reduction roadmap

| Unit: Megaton CO2e | 2018 | 2019 | 2020 | 2021 | 2022 | 2023 | 2024 | 2025 | 2026 | 2027 | 2028 | 2029 | 2030 | 2031 | 2032 | 2033 | 2034 | 2035 | 2036 | 2037 | 2038 | 2039 | 2040 | 2041 | 2042 | 2043 | 2044 | 2045 | 2046 | 2047 | 2048 | 2049 | 2050 |
|---|---|---|---|---|---|---|---|---|---|---|---|---|---|---|---|---|---|---|---|---|---|---|---|---|---|---|---|---|---|---|---|---|---|
| Actual GHG emissions | 296.32 | 287.76 | | | | | | | | | | | | | | | | | | | | | | | | | | | | | | | |
| Historical Data (TWh) Energy Bureau 2018 | 274.03 | | | | | | | | | | | | | | | | | | | | | | | | | | | | | | | | 124.0 |
| Target | 17.40.89 | 17.75.6 | 170.25 | 222.26 | 266.48 | 255.25 | 265.55 | 305.05 | | 255.47 | | | | | | | | | | | | | | | | | | | | | | | |

Notes
[illegible notes]

Sources
[illegible]

| Historical compound growth | Rate | GDP at constant prices (NT$ bn) | GDP GHG emission intensity (kt CO2e/NT$ bn) | Population |
|---|---|---|---|---|
| Scenario 1 - Historical Data (2015-2019) | | 2.4% | 0.75 | 0.17% |
| Scenario 2 - Historical Data (2010-2019) | | 3.0% | 0.73 | 0.35% |
| Scenario 3 - Current trends (applied to 2020 over average 2010-2019) | | 2.4% | 0.73 | 0.05% |

This data is from IP #1
[illegible]
CAGR of emission 2.4%
CAGR of Population 0.35%
GDP at constant prices (2010-2019): 3.0%






\end{document}